\documentclass[sigconf,nonacm]{acmart}

\usepackage{booktabs} 
\usepackage{url}
\usepackage{multirow}
\usepackage{array}
\usepackage{color}
\usepackage{xspace}
\usepackage{pifont}
\usepackage{subcaption}
\usepackage{caption}
\usepackage{enumitem}
\usepackage{listings}
\usepackage{tabularx}
\usepackage{wasysym}
\usepackage{multirow}
\usepackage{tabulary} 
\usepackage[most]{tcolorbox}
\usepackage{xcolor}
\usepackage{graphicx}
\usepackage{hyperref}
\usepackage{svg}
\usepackage{fancyhdr}
 
\lstdefinelanguage{json}{
    morestring=[b]",
    morecomment=[l]{//},
    stringstyle=\color{black},
    literate=
     *{:}{{{\color{black}:}}}{1}
      {,}{{{\color{black},}}}{1}
      {[}{{{\color{black}[}}}{1}
      {]}{{{\color{black}]}}}{1}
      {\{}{{{\color{black}\{}}}{1}
      {\}}{{{\color{black}\}}}}{1}
}

\newcolumntype{L}[1]{>{\raggedright\let\newline\\\arraybackslash}p{#1}} 
\newcolumntype{C}[1]{>{\centering\let\newline\\\arraybackslash}p{#1}} 
\newcolumntype{M}[1]{>{\centering\arraybackslash}m{#1}}
\newcolumntype{R}[1]{>{\raggedleft\let\newline\\\arraybackslash}p{#1}} 
\newcommand{\citesec}[1]{Section~\ref{sec:#1}}
\newcommand{\citefig}[1]{Fig.~\ref{fig:#1}}

\newcommand{\citelisting}[1]{{Listing}~\ref{lst:#1}}
\newcommand{\citetable}[1]{{Table}~\ref{tab:#1}}

\newcommand{\etal}[1][]{%
\ifthenelse{\equal{#1}{}}{\emph{et~al.}\xspace}{\emph{et~al.}~\cite{#1}\xspace}%
}

\newcommand{\fsc}{FSpec\xspace}

\definecolor{alizarin}{rgb}{0.82, 0.1, 0.26}
\newcommand{\stepdesc}[1]{\noindent\textbf{#1}}

\newcommand{\mdstyle}[1]{\texttt{{#1}}}

\newcommand{\cmark}{\ding{51}} % check mark
\newcommand{\xmark}{\textbf{\textsf{X}}} % cross mark}
\newcommand{\failid}[1]{\hypertarget{fail:#1}{#1}}
\newcommand{\failref}[1]{\hyperlink{fail:#1}{#1}}
\expandafter\def\expandafter\normalsize\expandafter{%
    \normalsize%
    \setlength\abovedisplayskip{0pt}%
    \setlength\belowdisplayskip{8pt}%
    \setlength\abovedisplayshortskip{-8pt}%
    \setlength\belowdisplayshortskip{2pt}%
}

\newtcolorbox[auto counter]{reqbox}[1]{title={\bfseries #1},
  coltitle=white, 
  leftrule=0.25mm,
  rightrule=0.25mm,
  bottomrule=0.25mm,
  toprule=0.25mm,
  colframe=white!45!black, boxsep=2pt,left=4pt,right=4pt,top=3pt,bottom=3pt,  sharpish corners}
\AtBeginDocument{%
  }

\setcopyright{acmlicensed}
\copyrightyear{2018}
\acmYear{2018}
\acmDOI{XXXXXXX.XXXXXXX}
\acmConference[ICSE 2026]{International Conference on Software Engineering}{April 12-18,
  2026}{Brazil}
\acmISBN{978-1-4503-XXXX-X/18/06}

\newcommand{\fwname}{{\it S\MakeLowercase{a}FUZZ}\xspace}
\usepackage{xspace}
\DeclareRobustCommand{\fwnamebold}{\textbf{\textit{S\MakeLowercase{a}FUZZ}}\xspace}

\newcommand{\droneresponse}{\textit{Drone Response}\xspace}

\definecolor{rev2}{RGB}{159, 43, 104} 
\newcommand{\suppurl}{https://github.com/SAREC-Lab/saFUZZ_ICSE26}

\begin{document}
%% pagenumber for the preprint
\setlength{\footskip}{30pt}
\pagestyle{fancy}
\fancyfoot[R]{\hspace{15mm}\thepage}
\title[Uncovering Failures in Cyber-Physical Systems]{Uncovering Failures in Cyber-Physical System State Transitions: \\A Fuzzing-Based Approach Applied to sUAS}

\author{Theodore Chambers}
\email{tchambe2@nd.edu}
\orcid{0003-2923-3230}
\author{Arturo Miguel Russell Bernal}
\email{arussel8@nd.edu}
\orcid{0009-2902-5766}
\affiliation{%
  \institution{University of Notre Dame}
  \city{Notre Dame}
  \state{Indiana}
  \country{USA}
}
% \author{Arturo Miguel Russell Bernal}
% \email{arussel8@nd.edu}
% \orcid{0009-2902-5766}
% \affiliation{%
%   \institution{University of Notre Dame}
%   \city{Notre Dame}
%   \state{Indiana}
%   \country{USA}
% }
\author{Michael Vierhauser}
\email{Michael.Vierhauser@uibk.ac.at}
\orcid{0003-2672-9230}
\affiliation{%
  \institution{University of Innsbruck}
  \city{Innsbruck}
  \country{Austria}
}
\author{Jane Cleland-Huang}
\email{janehuang@nd.edu}
\orcid{0001-9436-5606}
\affiliation{%
  \institution{University of Notre Dame}
  \city{Notre Dame}
  \state{Indiana}
  \country{USA}
}

\renewcommand{\shortauthors}{T. Chambers, A. Russell, M. Vierhauser, J. Cleland-Huang}

\begin{abstract}
The increasing deployment of small Uncrewed Aerial Systems (sUAS) in diverse and often safety-critical environments demands rigorous validation of onboard decision logic under various conditions.  In this paper, we present SaFUZZ, a state-aware fuzzing pipeline that validates core behavior associated with state transitions, automated failsafes, and human operator interactions in sUAS applications operating under various timing conditions and environmental disturbances.  We create fuzzing specifications to detect behavioral deviations, and then dynamically generate associated Fault Trees to visualize states, modes, and environmental factors that contribute to the failure, thereby helping project stakeholders to analyze the failure and identify its root causes. We validated SaFUZZ against a real-world sUAS system and were able to identify several points of failure not previously detected by the system's development team. The fuzzing was conducted in a high-fidelity simulation environment, and outcomes were validated on physical sUAS in a real-world field testing setting. The findings from the study demonstrated SaFUZZ's ability to provide a practical and scalable approach to uncovering diverse state transition failures in a real-world sUAS application.
\end{abstract}


\maketitle
\begingroup
\renewcommand{\thefootnote}{}%
\footnotetext{This is the author-accepted manuscript of a paper accepted for publication at ICSE 2026. The final published version may differ.}

\addtocounter{footnote}{0}%
\endgroup

% OLD VERSION IS AT THE BOTTOM

\section{Introduction}
\label{sec:intro}

Cyber-Physical Systems (CPS) are increasingly deployed across various domains, including transportation, healthcare, and robotics~\cite{falayi2025edge,kumar2024implementation,campusano2021towards,haque2014review}, where they operate in the physical world, under potentially dynamic and uncertain environmental conditions. To manage inherent complexity, CPS are often built using state machines that manage mode transitions and help ensure predictable operations~\cite{bartocci2021cpsdebug,amir2017hybrid,smyczynski2017autonomous}. Small Uncrewed Aerial Systems (sUAS) represent a rapidly growing class of CPS with wide-ranging applications including search-and-rescue ~\cite{Adams2009,frasheri2018adaptive}, environmental monitoring, infrastructure inspection~\cite{mcaree2016model}, surveillance, and disaster response~\cite{anand2023drones}. Their state-machines enable mission-level capabilities while leveraging lower-level services of the flight controller (aka an \emph{autopilot})~\cite{PX4,ardupilot}. These two layers interact via protocol-level interfaces such as MAVLink~\cite{mavlink}, creating implicit dependencies and complex interactions. Adding to this complexity, human operators can override automated tasks from a remote handheld controller (RC), for example, by issuing mode change requests to trigger failsafe mechanisms, such as Return-to-Launch (RTL). System behavior emerges from the composition of multiple state machines at various levels of operation, where application logic, flight controller firmware, and human input can all influence system states. Each level makes distinct assumptions, has unique timing constraints, and provides its own failure modes. As these interacting state machines grow in complexity, they become difficult to validate exhaustively, increasing the risk of latent design flaws, unintended interactions, or inadequate handling of edge cases.

Several previous drone incidents have highlighted problems associated with \emph{complex state behavior}, \emph{failsafe logic}, and \emph{human interactions}. For example, in 2022, a commercial drone pilot, reportedly lost control of his sUAS as a result of connection issues following a mode switch, and subsequently crashed into a manned aircraft deployed on a wildfire suppression mission~\cite{skydance_incident}. Similarly, in a 2020 incident, an sUAS experienced unreliable GPS/compass signals that triggered an automatic transition out of position-control into a degraded attitude-hold mode; the operator was initially unaware of the mode change, leading to loss of control and a crash \cite{Hambling2020GPSInterferenceDroneCrash}. 
In both cases, contributing factors included insufficient visibility into internal state transitions, failed mode changes, and delayed or ineffective human intervention. Both incidents underscore the need to systematically validate the behavior of application-level logic and flight controller firmware, including their composition and response under adverse conditions such as signal loss, sensor failure, or operator confusion. While the use of formal verification methods can provide strong behavioral guarantees, applying them in this type of complex, evolving software system, is currently limited to well-scoped components and remains impractical for full-system validation~\cite{bolton2013using,anto2023}. 

Our work addresses this gap through semantic-level fuzz testing of sUAS state machines. We propose \fwname, a novel fuzz-based framework for validating cross-layer and multi-agent interactions, with a particular focus on hazards emerging around mode transitions, failsafe behaviors, and control handoff. In contrast to code-based fuzzing approaches~{\cite{lemieux2018fairfuzz,visser2020coastal,holler2012fuzzing,bekrar2011finding}}, rather than mutating low-level inputs, \fwname systematically generates semantically meaningful event sequences drawn from realistic missions, incorporating variations in timing and environmental conditions that may influence system behavior. This approach explores the impact of both expected and unexpected events, enabling the discovery of subtle faults introduced by missing transitions, incomplete configurations, unexpected timing interactions, or misaligned assumptions between the application, flight controllers, and human operators under realistic real-world conditions.  The work makes the following contributions, with applications to both research and practice. \vspace{-2pt}
\begin{itemize}[leftmargin=*,itemsep=0.3em]
    \item[-]We present \fwname, a strategic fuzz-testing framework that incorporates behavioral semantics to analyze system-level state transitions and event sequences, enabling the identification of  transition hazards and unsafe state compositions (cf.~\citesec{pipeline}).

    \item[-]We provide a reusable experimental method for inducing and analyzing transition-related hazards arising from interactions among application-level and flight-controller state machines, including mode transitions, failsafes, and control handoff. \fwname enables systematic exploration of unsafe state compositions and helps reveal conditions under which common verification techniques may fail (cf.~\citesec{pipeline}).

    \item[-] We demonstrate the applicability of \fwname on a representative autonomous sUAS platform, using hazard-driven automated test generation to reveal previously undocumented vulnerabilities in failsafe logic and cross-layer interactions (cf.~\citesec{validation}).

    \item[-] We provide structured research artifacts, including decision-tree oracles for classifying test outcomes and automatically generated Fault Trees, as supplemental material. \emph{See supplemental material at \url{\suppurl}.}
\end{itemize}

The remainder of the paper is structured as follows.
 In \citesec{background}, we provide a brief introduction to sUAS systems and flight controllers and motivating examples for our work. Then, in \citesec{pipeline}, we introduce our \fwname framework and the steps of the automated pipeline. In~\citesec{validation}, we describe our evaluation setup for addressing feasibility, failure detection, and testing automation, and in \citesec{application}, we report on the results of our three research questions. Finally, we discuss threats to validity in~\citesec{threats}, related work in~\citesec{relwork}, and conclusions in \citesec{conclusions}.

\section{sUAS Layered Architectures}
\label{sec:background}

Modern sUAS systems typically follow a layered architecture that separates real-time control, flight behavior, communication, and mission-level decision-making. Each layer manages specific responsibilities, with state machines at different levels reflecting different scopes of autonomy and abstraction. At the lowest level, the flight controller directly interfaces with sensors and actuators and runs tightly coupled control loops such as rate and attitude stabilization.

PX4~\cite{PX4} and ArduPilot~\cite{ardupilot} are two of the most widely used open-source flight control software platforms for drones, supporting a wide range of aerial (fixed-wing, multirotor, VTOL) and terrestrial (UGV) vehicles. Both share similar capabilities and enforce hard safety constraints like preflight arming checks and emergency disarming, and they operate with strict real-time guarantees. This layer executes low-level control commands in real time.
For example, as part of the autopilot of the flight controller, PX4 defines a finite-state machine for managing high-level flight modes, such as \texttt{STABILIZED}, \texttt{POSCTL}, \texttt{OFFBOARD}, \texttt{RTL}, and \texttt{LAND}~\cite{px4modes}. Each mode enables or restricts certain control inputs, and transitions are based on operator commands, autonomous triggers, or failsafe events. This state machine enforces operational to ensure valid and safe transitions.

\begin{figure}[t]
\centering

\includegraphics[width=\linewidth]{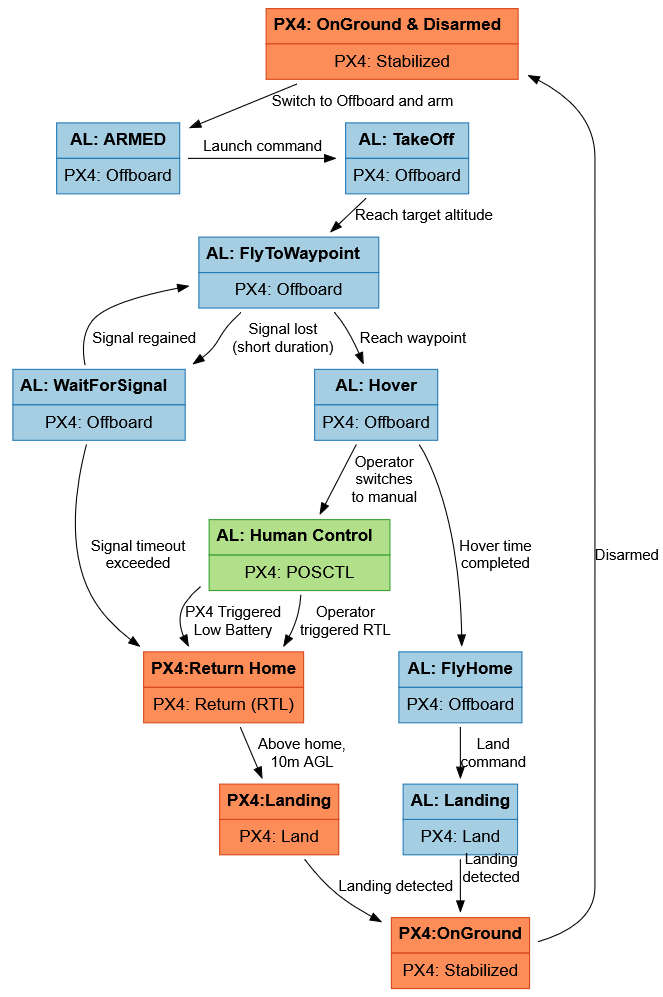}
\caption{Application-level states and lower-level PX4 modes for a simple mission. Colors indicate the controlling entity: Application (Blue), PX4 (Orange), Human (Green).}
\label{fig:merged_states}
\Description{A small subset of the State Machine for a multi-level sUAS application showing application-level states}
\vspace{-1em}
\end{figure}

\begin{figure*}[th!]
\centering
\includegraphics[width=0.999\textwidth]{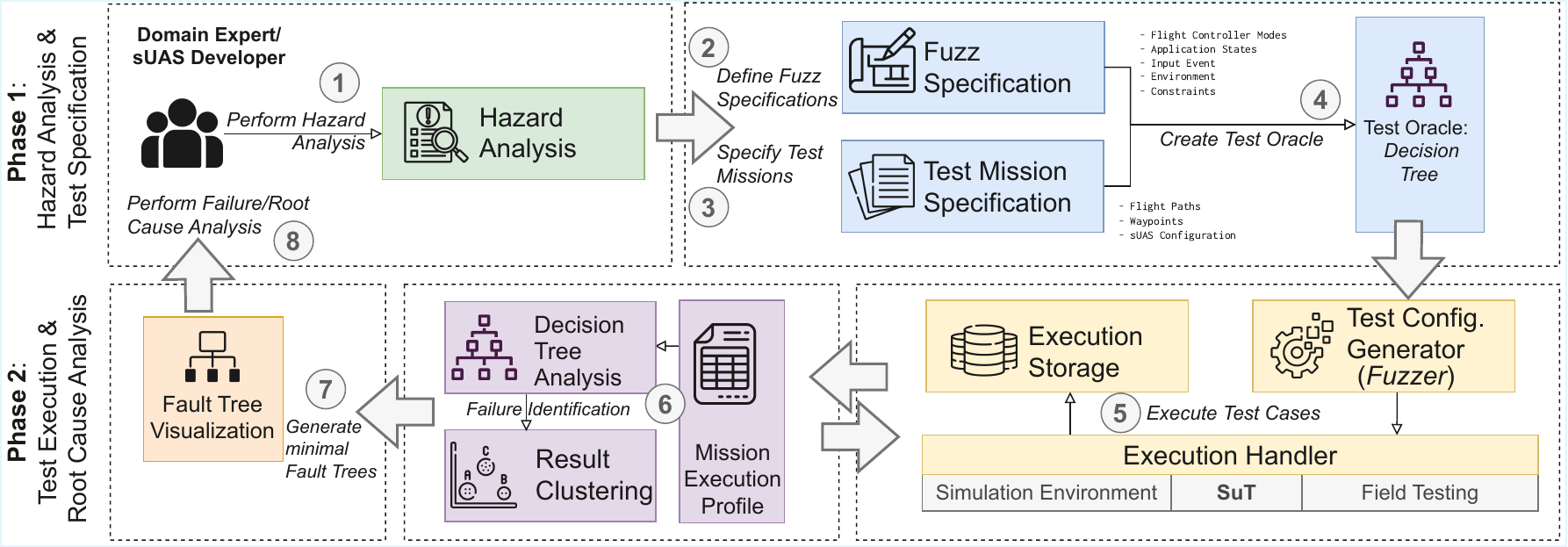}
\caption{High-Level overview of \fwnamebold showing preliminary setup (Phase 1) and the automated pipeline (Phase 2).}
%\vspace{-12pt}
\Description{High-Level overview of the presented framework with the core components and two phases.}
\label{fig:overview}
\end{figure*}

sUAS applications typically use the MAVLink~\cite{mavlink} communication protocol to interact with the autopilot, operating in \mdstyle{OFFBOARD} (PX4) mode to stream high-frequency position, velocity, or attitude setpoints to the flight controller. This supports fine-grained maneuvers for collision avoidance and path following. In addition, applications commonly implement mission-level state machines that depend on, and compose with, the flight controller’s internal state machine. This is illustrated in \citefig{merged_states}, which shows the interplay between PX4 and application-layer states during a simple mission. Each node depicts an application-level state (e.g., FlyToWaypoint) and a PX4 mode (e.g., \mdstyle{OFFBOARD}). Blue states are controlled at the application level, orange states by PX4, and green ones by a human operator who has intervened in the mission by switching to a manual flight mode (\mdstyle{POSCTL}).  While not depicted in this diagram, the application layer may also implement its own failsafes. For example, developers might configure the PX4-level loss-of-signal failsafe to trigger after 60 seconds, while setting an application-level failsafe to trigger after only 20 seconds of lost signal, thereby providing the opportunity to recover or adjust the mission before the PX4-level failsafe activates.

These types of interactions between application-level and autopilot state machines can be quite complex, with errors arising at multiple levels~\cite{siewert2019fail}. State-based failures in state configurations and transitions can trigger unintended behaviors, such as aborted missions~\cite{rocamora2024behavior}. Failures can also arise when failsafe mechanisms, designed to handle contingencies (e.g., compass interference or loss-of-signal), behave incorrectly introducing new forms of failures.  Human operators may also contribute to failures by misinterpreting events, issuing delayed responses, or providing incorrect commands that result in unsafe interventions. Finally, feature-interaction failures can occur when independently correct behaviors interact in unexpected ways. Because these interactions often arise during contingency or time-critical mission states, they can produce sudden, irrecoverable behaviors that pose heightened safety risks and are particularly difficult to anticipate or reproduce during testing.

\section{The \fwnamebold Framework}
\label{sec:pipeline}

To address these challenges, we present \fwname, our automated fuzz testing pipeline for validating state behavior and transitions in sUAS applications. We followed the Design Science methodology~\cite{wieringa2014design} to develop \fwname through iterative cycles of problem analysis, design, simulation, implementation, and evaluation. 
As depicted in \citefig{overview}, \fwname includes two main phases and eight steps. In Phase 1, project stakeholders perform hazard analysis (Step 1), and construct Fuzz Specifications (Steps 2-4) which serve as the basis for subsequent tests. In Phase 2, the tests are automatically executed and analyzed, and Fault Trees are dynamically generated for each failed test (Steps 5-7). These Fault Trees are then presented to project stakeholders to support failure analysis (Step 8). Each step, including its associated artifacts and activities, is now described in more detail.

\begin{figure*}[]
    \centering
    \includegraphics[width=0.99\textwidth]{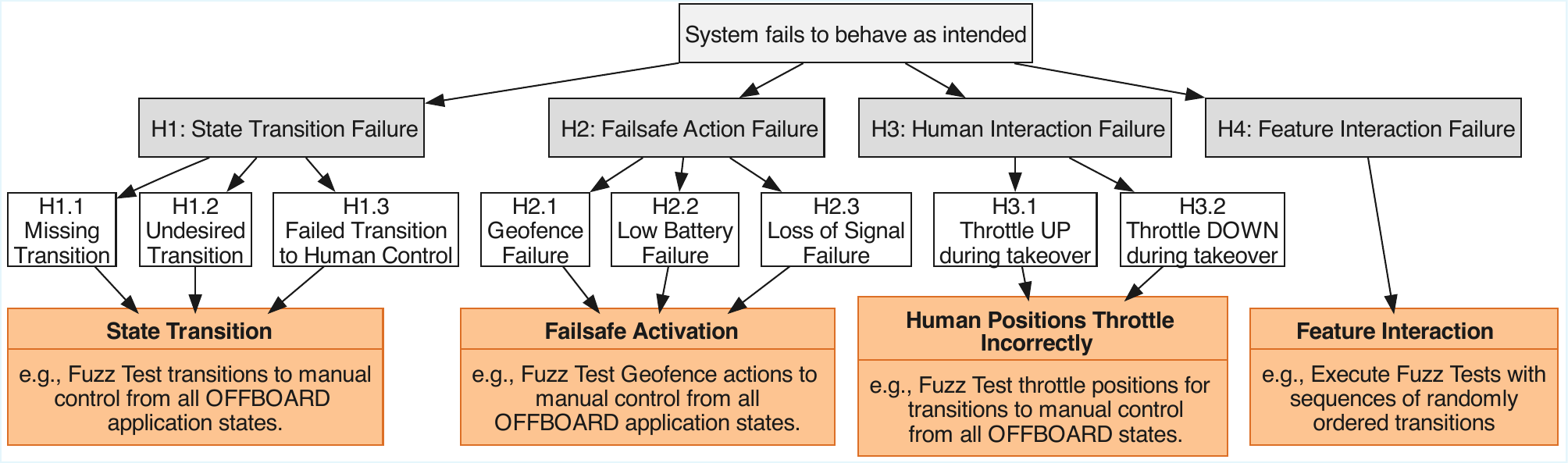}
    \caption{Partial hazard tree illustrating four representative categories of sUAS failures: state transition failures (H1), failsafe activation failures (H2), human interactions, illustrated here as incorrect positioning of the throttle upon human takeover (H3), and feature interaction errors (H4). Annotated nodes show example semantic fuzz tests targeting each category. The types of hazards shown in this tree are not exhaustive.}
    \Description{Partial hazard tree illustrating representative classes of sUAS failures}
    \label{fig:hazards}
    %\vspace{-12pt}
\end{figure*}

\subsection{Phase 1: Hazard Analysis \& Specification}

\stepdesc{Step 1 -- Hazard Analysis:} 
We adopted a hazard analysis approach to guide the focus of the testing process~\cite{golabi2022towards,plioutsias2018hazard}. 
\citefig{hazards} shows a partial hazard tree exploring four types of failure related to state transitions, failsafe actions, and throttle positions. The first type of hazard is fundamental to any state machine (\emph{H1}), as it is essential to validate that all transitions can execute correctly. The second addresses the safety-critical need for failsafe mechanisms (\emph{H2}), able to handle common faults such as geofence encroachment, low battery, and loss-of-signal. The third represents the family of hazards associated with human errors. Here we use the example of positioning the throttle on the RC Transmitter into potentially dangerous positions (\emph{H3}). 
Finally, we explore Feature Interaction Errors (\emph{H4}) through a broader set of fuzz tests. 
\vspace{4pt}
\stepdesc{Step 2 -- Fuzz Specifications:}
Based on these identified hazards, domain experts define corresponding {\it Fuzz Specifications} (\fsc). Each Fuzz specification defines a test space by specifying possible combinations of flight-controller modes, application states, environmental factors, injected actions, and timing. 
\vspace{4pt}
\definecolor{lightgray}{rgb}{0.2,0.2,0.2}
\begin{lstlisting}[
  caption={Fuzz Specification 1 (\fsc-1): Mode transitions during autonomous flight. The Specification addresses Hazard H1.},
  label={lst:fuzz-scenario-1},
  frame=single,
  rulecolor=\color{lightgray},
  basicstyle=\ttfamily\footnotesize,
]
{
  "FROM_PX4_modes": ["OFFBOARD", "LAND"],
  "FROM_APP_states": ["TAKEOFF", "FLYING_TO_WAYPOINT", "HOVERING", "LANDING", "DISARMING"],
  "RC_INPUT_EVENTS": ["ALTCTL", "POSCTL", "STABILIZED"],
  "ENVIRONMENT": {
    "transition_delay": {
        "bands": {
            "short": { "min": 50,  "max": 200  },
            "medium":{ "min": 200, "max": 600  },
            "long":  { "min": 600, "max": 1200 }
    }    }
    "throttle": ["mid"],    "geofence": ["none"],
    "wind": ["none"],       "GPS": ["none"],
    "COMPASS_INTERFERENCE": ["none"]
  },
  "MISSION_CONTEXT": ["Flight plan A"],
  "CONSTRAINTS": {
    "REQUIRES_PX4_MODE": {
      "OFFBOARD": ["TAKEOFF", "FLYING_TO_WAYPOINT*", "HOVERING"],
      "LAND": ["LANDING", "DISARMING"]
}}}
\end{lstlisting}

\noindent The example in \citelisting{fuzz-scenario-1} represents Hazard H1 by defining RC input events, timing variations, throttle position, and other environmental factors within valid PX4–application state combinations, allowing \fwname to check whether state transitions occur as expected\footnote{\label{fn:supp}The full vocabulary for the Fuzz Specifications can be found in supplemental material at \url{\suppurl}.}.
\vspace{3pt}

\stepdesc{Step 3 -- Specify Test Mission:}
A mission provides the execution context for tests generated from a Fuzz Specification. It defines the flight path to be flown and maneuvers to be performed, ensuring that the sUAS will naturally progress through the states and modes required to reach the targeted test context. When that context is observed (i.e., the current state, operational mode, and environmental conditions match the specified criteria), a timer is activated and the designated mode transition is injected with the appropriate delay. This combination of state, mode, environmental context, timing delay, and injected mode transition defines the specific test instance generated by the Fuzz Specification and tested during execution.
In other words, if the test calls for triggering an event from FlyToWaypoint/\mdstyle{OFFBOARD}, then this state/mode combination must occur during the flight for the test to be valid. Similarly, tests associated with environmental factors, such as geofence actions,  must also include a flight path that crosses or approaches a geofence.\vspace{3pt}

\stepdesc{Step 4 -- Create a Test Oracle:} To evaluate the outcome of each test, \fwname requires a test oracle. While prior work has validated sUAS test outcomes based on simple logic, such as whether the sUAS completes its mission within a fixed time and adheres closely to the planned flight path~\cite{DBLP:conf/seams/PurandareSICC23, chambers2024hifuzz}, this is insufficient for validating correct behavior of state transitions and mode changes. For example, if a test validates that a \mdstyle{POSCTL} mode change is activated during flight, then it would be an error if, in fact, the sUAS  completed the mission as planned instead of exiting the mission and entering a hover state (as expected in \mdstyle{POSCTL}). 
Therefore, in order to fully automate the pipeline, the test oracle is constructed in the form of a decision tree that provides a guided process for differentiating between three different types of outcomes. These outcomes include (a) \emph{invalid tests}, which occur when the test’s targeted conditions are not met (i.e., a fault in the test case itself) or an excessive timing delay causes the test to be executed in the wrong context, (b) \emph{passing tests}, in which the test executes as planned and all success criteria are met, and (c) \emph{failing tests}, in which the test executes as planned but at least one success criterion is not achieved.

As illustrated in Figure~\citefig{decision-tree}, the decision tree encodes system-level post-conditions for state transitions, mode changes, and failsafe behavior, and uses these semantics to differentiate among invalid, passing, and failing tests. Because these test outcomes are defined at the SuT level, the same decision tree oracle applies uniformly across all tests generated from a Fuzz Specification.  In addition to constructing the decision tree, we identify any data that it requires as inputs during the analysis process, and ensure that the system is instrumented to collect and deliver this data. Designing, testing, and refining, the decision tree is time-consuming, and requires significant domain expertise. 
%\vspace{-1pt} %\jch{TODO:Describe in pipeline}

\begin{figure}[t!]
  \centering
    \includegraphics[width=\linewidth, trim=0 0 45 0, clip]{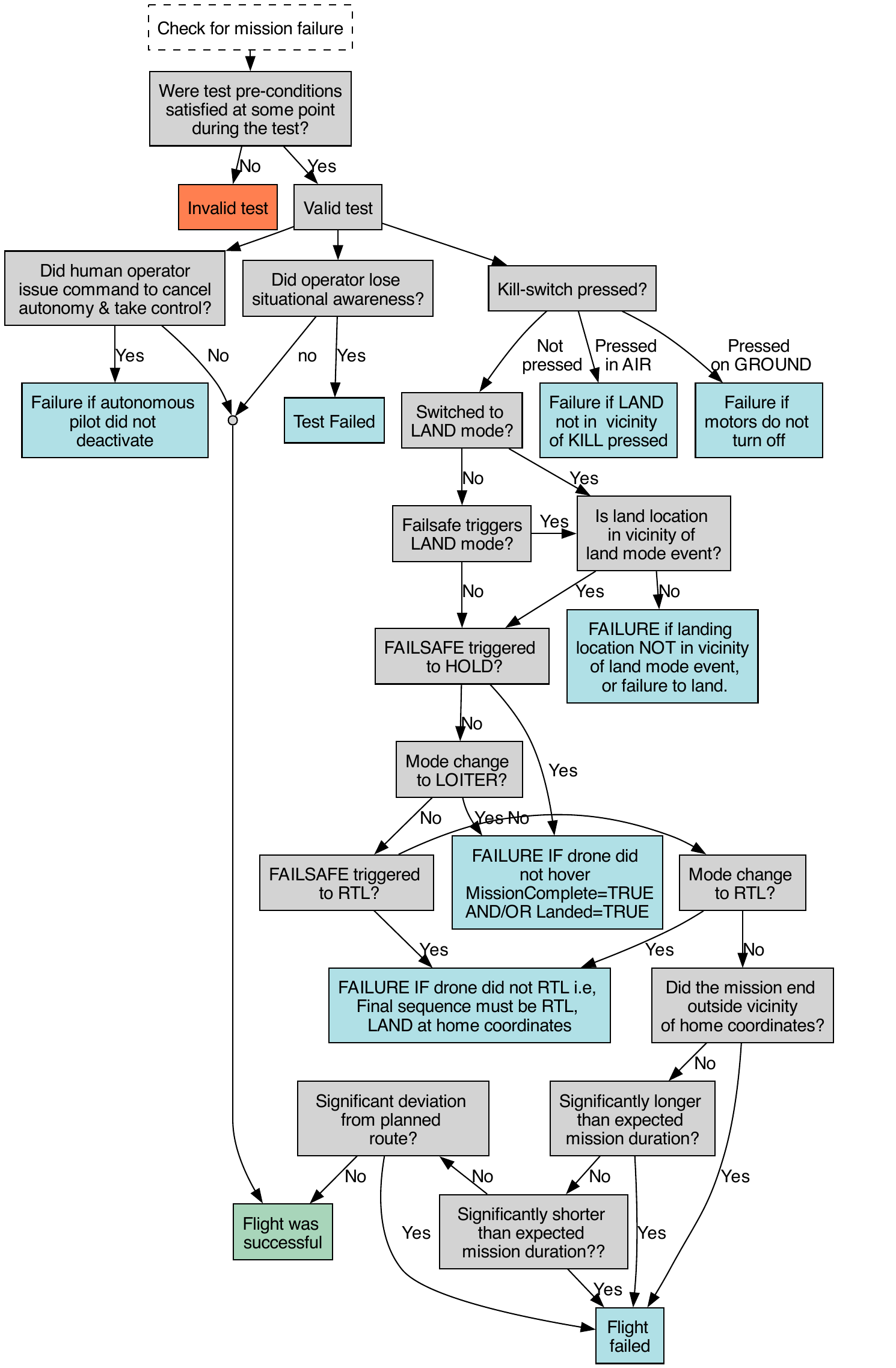}
  % \includesvg[width=\linewidth, inkscape=false]{figures/short-decision-tree.svg}
  \caption{Decision Tree classifier for labeling fuzz‐test outcomes.  Nodes check mission‐failure predicates and human‐interaction conditions; assigns \emph{Success}, \emph{Failure}, or \emph{Invalid}. }
  \label{fig:decision-tree}
  \Description{Decision Tree classifier for labeling fuzz‐test outcomes.}
  \vspace{-12pt}
\end{figure}
%All raw telemetry, state transitions, exception flags, and failure indicators are persisted for comprehensive offline analysis.
\vspace{-12pt}
\subsection{Phase 2: Test Execution \& Analysis}
The second phase of \fwname automates the fuzz testing process. We describe it here using mode names from the PX4  autopilot.\vspace{2pt}

\stepdesc{Step 5 -- Test Execution:}
\fwname  accepts a Fuzz Specification as input and iterates through hundreds (or thousands) of individual test cases. Each test includes the following steps:
\begin{enumerate}[leftmargin=.9em,itemsep=0.2em]
\item[-]\textbf{Environment Setup:~}
    Environmental variables defined in the specification are configured before test execution. For example, a geo\-fence is setup, and/or wind parameters,GPS, and compass interference initialized in the simulation environment.
    \item[-]\textbf{State Selection:~}
    A valid PX4 mode and corresponding application-level state are selected in accordance with the constraints defined in the test specification. 
    \item[-]\textbf{Action Injection:~}
    A control action (e.g., \mdstyle{ALTCTL}, \mdstyle{POSCTL}) is chosen at random and scheduled to be applied during the selected state. 

    \item[-]\textbf{Timing Configuration:~}
    A transition delay band (\textit{short}, \textit{medium}, or \textit{long}) is selected, and delay time (in ms) is sampled from the corresponding range and applied before the action is dispatched.

    \item[-]\textbf{System Configuration:~}
    The sUAS is initialized using the defined \mdstyle{MISSION\_CONTEXT} (e.g., a predefined flight plan), and any mode-specific settings (e.g., geofence behaviors) are applied.

    \item[-]\textbf{Test Execution:~}
    The test flight is launched in a high-fidelity simulation environment (e.g., Gazebo~\cite{gazebo}), and the system is monitored to detect the \emph{first occurrence} of the targeted combination of mode, state, and environment factors, at which point the control action is injected with the configured timing delay.

    \item[-]\textbf{Monitoring and Outcome Logging:~}
    During test execution, telemetry and behavior logs are captured as specified by the test oracle and stored in the \emph{Execution Storage} for later analysis.     \item[-]\textbf{Environment reset:~}The entire environment is dynamically reset between each test case.
\end{enumerate}

\stepdesc{Step 6 --  Failure Identification:} 
Each test execution produces a corresponding JSON  profile documenting the test outcome (as illustrated in supplemental materials). Failures are identified through a structured process combining decision-tree classification and clustering-based anomaly detection as follows:
\begin{enumerate}[leftmargin=.9em]
    \item[-]\textbf{Failure Case Identification:~} The outcome of each individual test case is automatically analyzed using the decision tree logic (cf.~\citefig{decision-tree}). Outcomes landing on blue nodes are tagged as \mdstyle{FAILED}.

    \item[-]\textbf{Test Selection:~} Due to the fuzzing process, \fwname executes many similar tests with closely related results. Therefore, for each Fuzz Specification \(\mathcal{FS}\) we selected the FAILED tests, and apply \(K\)-means clustering algorithm to their feature vectors, using the elbow method to determine K \cite{cui2020elbow}. Each test $T_i$ is labeled with an outcome $y_i$, where $y_i = 1$ indicates a failed test that has violated expected behavior. Continuous parameters are normalized, and categorical parameters are one-hot encoded to prepare the data for clustering.     
    The homogeneity of each cluster $C_j$ is defined by the \textit{within-cluster sum of squares} (WCSS):
    
    \[
    \text{WCSS}_j = \sum_{\mathbf{x}_i \in C_j} \|\mathbf{x}_i - \boldsymbol{\mu}_j\|^2,
    \]
    where \(\boldsymbol{\mu}_j\) is the centroid of cluster $C_j$. This metric measures the total squared Euclidean distance from all cluster points to the centroid. Following this, we select the test closest to the centroid, and the test farthest from the centroid from each cluster for initial next-step analysis. 

\end{enumerate}

\stepdesc{Step 7 -- Fault Tree Generation and Visualization:~} 
The previous clustering and analysis step identifies individual failures. However, the commonalities driving these failures are hard to analyze directly from the raw data. Therefore, \fwname performs a second round of highly focused fuzzing around the selected tests, using the test outcomes to generate Fault Trees that visualize each type of failure. The steps are as follows:

\begin{table}[t]
  \captionsetup{justification=centering}
  \centering
  \caption{Truth Table for \mdstyle{TAKEOFF} with \mdstyle{POSCTL} switch across varied timing intervals (cf. Failure F2).}
  \label{tab:truth-table-ex}
  \vspace{-10pt}
  \small
  \renewcommand{\arraystretch}{1.1}
  \setlength{\tabcolsep}{3pt}
  \begin{tabular}{|R{1.3cm}| R{1.65cm}| c| c| c|}
    \hline
    \textbf{App. State}   &  \textbf{Mode Switch}  &  \textbf{Interval (ms)}         &  \textbf{Status} &  \textbf{Failure Rate} \\
     \hline
    \mdstyle{TAKEOFF} & N/A     & N/A                    & 0      & 0\%     \\ \hline
    \mdstyle{TAKEOFF} & \mdstyle{POSCTL}  & 50–1000 (Short)      & 1      & 100\%   \\ \hline
    \mdstyle{TAKEOFF} & \mdstyle{POSCTL}  & 1000–5000 (Medium) & 1      & 65\%    \\ \hline
    \mdstyle{TAKEOFF} & \mdstyle{POSCTL}  & 5000–10000 (Long)  & 0      & 0\%     \\ \hline
  \end{tabular}
  \vspace{1ex}

  \footnotesize
  Failure Rate = percent of runs with at least one failure; Test Status: 1 = Fail, 0 = Pass. Rates based on 20 runs per interval.
    \vspace{-12pt}
\end{table}

\begin{table}[t]
  \centering
  \caption{Failure Breakdown when considering the current mode at the time the mode switch appeared (cf. Failure F2).}
  \label{tab:truth-table-medium}
    \vspace{-10pt}
  \small
  \renewcommand{\arraystretch}{1.2}
  \setlength{\tabcolsep}{4pt}
  \begin{tabular}{|r| r| r| c| c|}
       \hline
     \textbf{App. State} &  \textbf{Current Mode}                 &  \textbf{Mode Switch} &  \textbf{Status} &  \textbf{Failure Rate} \\
       \hline
    \mdstyle{TAKEOFF}       & STABILIZED\textsuperscript{\dag} & \mdstyle{POSCTL}      & 1      & 100\%        \\    \hline
    \mdstyle{TAKEOFF}       & OFFBOARD                      & \mdstyle{POSCTL}      & 0      & 0\%          \\
        \hline
  \end{tabular}

  \vspace{1ex}
  \footnotesize
  \textsuperscript{\dag} All medium-interval failures occurred when \mdstyle{POSCTL} was triggered during the standard autopilot STABILIZED mode.
  \vspace{-12pt}
\end{table}
\begin{enumerate}[leftmargin=.8em]
\item[-]\textbf{Execute additional Fuzz Tests:~} We execute additional fuzz tests focused around each selected failure case to cover valid predicate combinations at multiple timing intervals. 
\item[-]\textbf{Generate a Truth Table:~} The results from these tests are used to automatically populate a truth table characterizing each failure profile, where each row corresponds to a unique combination of conditions. For example, Table~\ref{tab:truth-table-ex} presents the truth table generated from 20 runs for a test where \mdstyle{POSCTL} was applied in the \mdstyle{TAKEOFF} state with varying timing intervals. 

Normal behavior of the \mdstyle{TAKEOFF} state begins in \mdstyle{STABILIZED} mode before transitioning to \mdstyle{OFFBOARD}. Our test results indicate that issuing the \mdstyle{POSCTL} mode change succeeds reliably only after the sUAS has entered \mdstyle{OFFBOARD} (as shown in Table~\ref{tab:truth-table-medium}) -- a behavior previously unknown to our team. By analyzing the timing delays in the truth table, we were able to pinpoint the root cause of this failure. 

 \item[-]\textbf{Generate a Fault Tree:~}
 Once the truth table is complete, we extract minimal cut sets of predicate conditions sufficient to cause failure, using an algorithm inspired by the Quine–Mc Cluskey Boolean minimization method~\cite{quine1952problem}. However, unlike Quine–McCluskey, which exhaustively finds all prime implicates across the entire input space, our approach operates directly on the subset of failing test cases and is restricted to only VALID test combinations imposed by the state machine.

\begin{table*}[]
\centering
\footnotesize
\caption{Summary of three Fuzz Specifications \fsc-1 to \fsc-3  used for validating \fwnamebold. Additional examples are provided as supplemental materials (see Footnote~\ref{fn:supp}).}
\label{tab:fuzz-scenarios}
\renewcommand{\arraystretch}{1.2}
\begin{tabular}{|L{1.95cm}|L{4.6cm}|L{4.2cm}|L{5.4cm}|}
\hline
\textbf{Fuzz Specification} & \textbf{\fsc-1} & \textbf{\fsc-2} & \textbf{\fsc-3} \\ \hline
\textbf{Overview}&Test human control across multiple states&Test Failsafe actions across two states&Test Failsafe actions triggered by geofence\\
\hline
\textbf{PX4 Modes} &
\mdstyle{OFFBOARD}, \mdstyle{LAND} &
\mdstyle{OFFBOARD} &
\mdstyle{OFFBOARD} \\
\hline
\textbf{Tested App States} &
\mdstyle{TAKEOFF}, \mdstyle{FLYING\_TO\_WAYPOINT}, \mdstyle{HOVERING}, \mdstyle{LANDING}, \mdstyle{DISARMING} &
\mdstyle{FLYING\_TO\_WAYPOINT}, \mdstyle{HOVERING}
&
\mdstyle{FLYING\_TO\_WAYPOINT} \\
\hline
\textbf{Tested Mode/Throttle activations} &
\emph{RC\_INPUT:} \mdstyle{ALTCTL}, \mdstyle{POSCTL}, \mdstyle{STABILIZED}, \mdstyle{THROTTLE\_TOGGLED} &
\emph{RC\_INPUT:} \mdstyle{AUTO.LOITER}, \mdstyle{AUTO.LAND}, \mdstyle{AUTO.RTL} &
\emph{GEOFENCE ACTIONS:} RTL (+LAND), LAND, WARNING\newline
\emph{RC\_INPUT\_EVENTS:} \mdstyle{ALTCTL}, \mdstyle{POSCTL}, \mdstyle{STABILIZED}, \mdstyle{OFFBOARD} \\
\hline
\textbf{Environment / Context} &
-- \emph{Delay:} short / medium / long \newline
-- \emph{Throttle:} mid / low \newline
-- \emph{Geofence:} none \newline
-- \emph{Wind, GPS, Compass:} none \newline
-- \emph{Context:} Flight plan A \newline
-- \emph{Constraints:} PX4 mode -- App state mapping &
-- \emph{Delay:} short / medium \newline
-- \emph{Throttle:} mid \newline
-- \emph{Geofence:} none \newline
-- \emph{Wind, GPS, Compass:} none | low / medium / high | low / medium / high \newline
-- \emph{Context:} Flight plan B &
-- \emph{Delay:} short / medium / long \newline
-- \emph{Throttle:} mid \newline
-- \emph{Geofence:} active | actions: \mdstyle{WARN}, \mdstyle{RETURN}, \mdstyle{LAND} \newline
-- \emph{Wind, GPS, Compass:} none \newline
-- \emph{Context:} Flight plan C \\
\hline
\textbf{Failures}
  & \failref{F2},  \failref{F8}, \failref{F9}
  & \failref{F1}, \failref{F5}, \failref{F6},  \failref{F10}, \failref{F11}
  & \failref{F3}, \failref{F4}, \failref{F7} \\
\hline
\end{tabular}
\end{table*}
 These trees represent the smallest conjunctions of state and environmental conditions likely to cause failures, effectively identifying minimal cut-sets of predicate combinations and pointing to root-cause conditions responsible for failures within the group. The Fault Tree associated with the truth table in Table~\ref{tab:truth-table-ex} is depicted in \citefig{ft2}.
 \end{enumerate}

\stepdesc{Step 8 -- Failure Analysis:~} 
Finally, the generated Fault Trees are made available to project stakeholders, such as developers and architects, for detailed analysis. There are two primary outcomes for each generated Fault Tree, including: 
\begin{itemize}[leftmargin=*]

\item[-] {\it The identified failure is a false positive}. This typically implies an error in the decision tree. For example, we encountered an error when the decision tree checked for transitions to Loiter (hovering), because PX4 transforms \mdstyle{AUTO.LOITER} to \mdstyle{POSCTL} as it exhibits similar behavior in rotorcraft. Without correctly accounting for this, \fwname raises an unexpected mode error. 

\item[-] {\it The Fault Tree depicts an actual bug}. This normally triggers a bug report or creation of a new issue, and may also trigger a deeper discussion recognizing that the observed behavior, while incorrect, is associated with a missing requirement. Additionally, it could trigger the definition and execution of new Fuzz Specifications that focus attention on the identified fault.
\end{itemize}
Although fuzz testing is uniquely positioned to reveal classes of failures that are difficult to detect by other methods, it offers no completeness guarantees, and undiscovered failures (i.e., false negatives) may still persist. Therefore, to close the loop, any unexpected behavior observed in simulation or during future field-deployments, informs the creation of new Fuzz Specifications, enabling targeted exploration of the conditions that triggered it.

\section{Evaluating \fwnamebold}
\label{sec:validation}
\begin{figure}[]
    \centering
    \includegraphics[
      width=0.95\linewidth,
      trim=0 1.5cm 0 1.5cm,
      clip
    ]{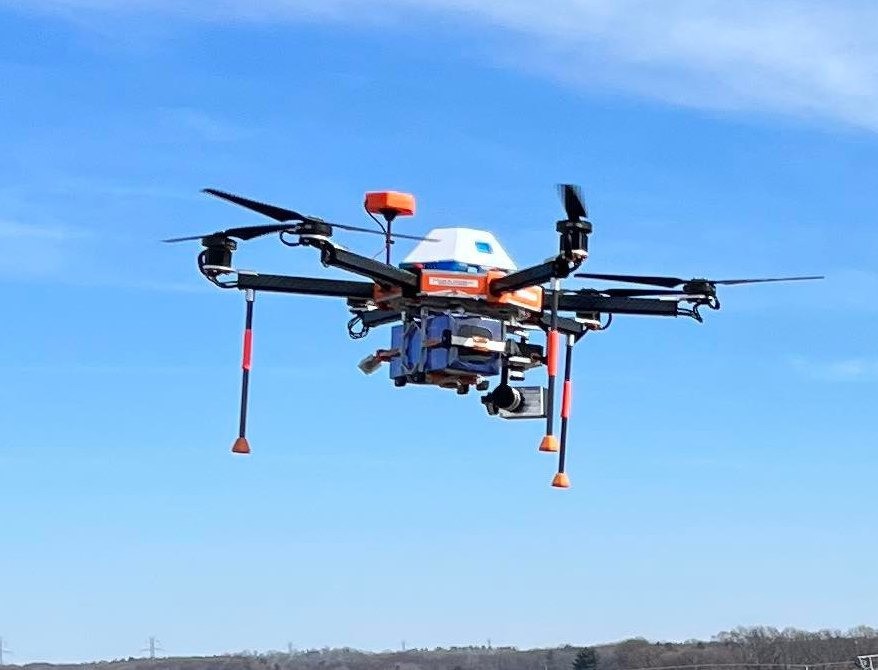}
    % \includegraphics[
    %   width=\linewidth,
    % ]{figures/sd2.jpg}
    \caption{One of the PX4-equipped hexacopters used in the field tests, running \textit{Drone Response} Autonomy software onboard a Jetson Xavier NX.}
    \label{fig:drones}
    % \centering
    % \includegraphics[
    %   width=0.95\linewidth,
    %   trim=0 1.0cm 0 1.0cm,
    %   clip
    % ]{figures/skydrones.png}
    % % \includegraphics[
    % %   width=\linewidth,
    % % ]{figures/sd2.jpg}
    % \caption{The \droneresponse software stack also supports swarming, as depicted here using two Inspired Flight 1200A drones.}
    % \label{fig:enter-label}
    \vspace{-16pt}
\end{figure}

To evaluate the effectiveness of \fwname, we used \droneresponse~\cite{droneresponse} as our system under test (SuT). \droneresponse has been built over the past eight years as part of our ongoing research program on autonomous sUAS  (e.g.,~\cite{cleland2018dronology,DBLP:conf/chi/AgrawalABCFHHTK20,DBLP:journals/taas/ClelandHuangCZCAV24,DBLP:journals/corr/abs-2505-08060,Chambers2025UTMOnEntry}). The platform has been developed by our research group, supported by a small professional software engineering team, typically consisting of two to three engineers at any given time. It represents a multi-sUAS management and control system, with a modular architecture comprising a configurable mission planner, a centralized ground control station, and onboard compute capability for real-time autonomy and perception. The system supports both PX4 and ArduPilot flight stacks, enabling integration with a range of airframes. Missions are orchestrated using a detailed operational state machine that governs critical mission stages, including pre-flight arming checks, autonomous takeoff, waypoint navigation across varied trajectories, stable hovering, landing, and safe disarming procedures after mission completion. 
We evaluate the effectiveness, scalability, and practical utility of \fwname using a January 2024 branch of \droneresponse as a real-world testbed, deployed on PX4-based sUAS as depicted in Figure \ref{fig:drones}, and structured around the following three research questions.\vspace{0.5em}

\begin{figure*}
\centering

\begin{subfigure}[t]{0.215\textwidth}
  \centering
  \raisebox{0pt}[\height][0pt]{%
    \includegraphics[width=\linewidth, trim=0 0 0 .1mm, clip]{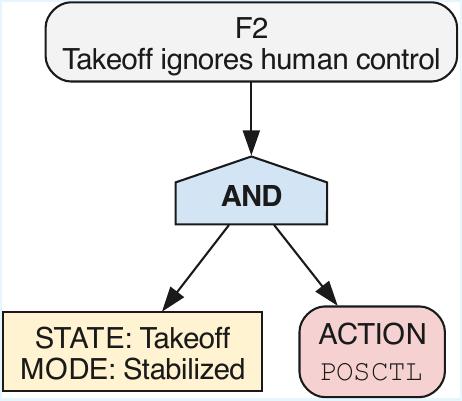}%
  }
  \caption{POSCTL was ignored during `stabilized' phase of Takeoff.}
  \label{fig:ft2}
\end{subfigure}\hfill
\begin{subfigure}[t]{0.32\textwidth}
  \centering
  \raisebox{3pt}[\height][0pt]{%
    \includegraphics[width=\linewidth, trim=0 0 0 .1mm, clip]{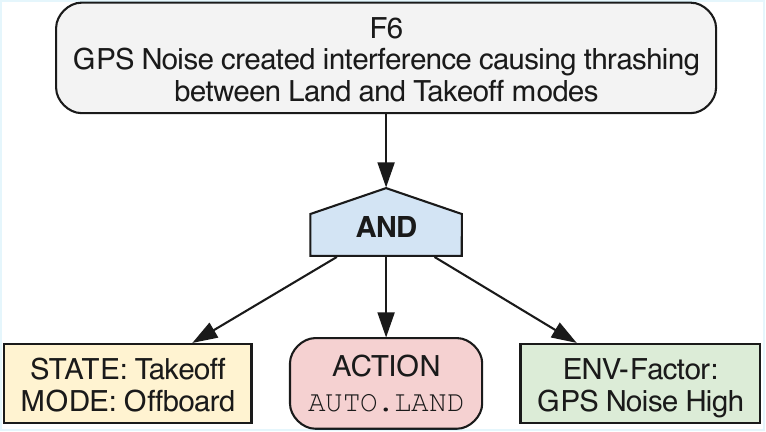}%
  }
  \caption{GPS noise, introduced into the test environment, caused unexpected mode changes.}
\end{subfigure}\hfill
\begin{subfigure}[t]{0.43\textwidth}
  \centering
  \raisebox{-1pt}[\height][0pt]{%
    \includegraphics[width=\linewidth]{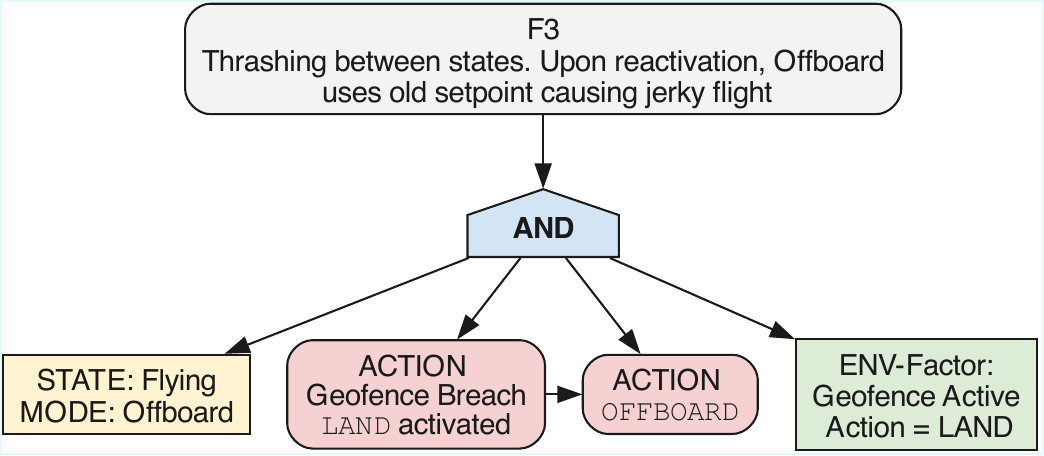}%
  }
  \caption{Thrashing occurred after a Geofence breach, when OFFBOARD was reactivated with a stale setpoint, causing jerky flight.}
\end{subfigure}

\vspace{-8pt}
\caption{\fwname identified eleven failure cases. Here we show the augmented Fault Trees for three different failure types. Each diagram highlights a root cause pattern observed during testing, and shows the current state (yellow), action(s) (pink), and environmental factors or configurations (green).}
\label{fig:failure_cases}
\vspace{-8pt}
\end{figure*}

\noindent\emph{\textbf{RQ1:}~To what extent can \fwname identify previously unknown behavioral failures in a real-world sUAS system?}~
This question examines whether our framework can effectively detect and categorize failures in the SuT. For each type of failure, we identify potential mitigations such as code modifications, requirements analysis, or updates to the decision tree.\vspace{0.2em}

\noindent\emph{\textbf{RQ2:}~How well do the transition-related errors detected by \fwname align with those identified by the development team over time?}
We conduct a detailed analysis of mode and state-related transition errors that existed in a January 2024 branch and compare the errors identified by executing \fwname versus those identified by the \droneresponse development team through the normal testing process over an 18 month period (Jan 2024-July2024).\vspace{0.2em}

\noindent\emph{\textbf{RQ3:}~To what extent are the failures identified by \fwname in simulation reproducible in real-world flight tests?} 
This final question assesses the correspondence between simulation-detected failures and their manifestation in physical flights. Where safe and feasible to do so, we replicate the test in the real world, to determine whether real-world behavior is consistent with \fwname's findings. 
\vspace{-6pt}

\subsection{\fwnamebold Experimental Prototype}
\fwname was developed for this research project as a fully executable prototype using Python 3.11.0, totaling roughly 4,000 lines of code. The prototype is organized into four modules, a \texttt{Fuzzer} component that creates the test configurations based on the Fuzz Specifications, an  \texttt{Executor} handling test execution in a simulator, a \texttt{Storage} component managing the simulation results, and an \texttt{Analyzer} component performing the clustering, decision tree analysis and Fault Tree generation. Each module was deployed as a Docker container, providing a high‐fidelity Gazebo‐based digital replication of the \droneresponse autonomy system, and ensuring that every test was executed in a clean, versioned environment, with automated teardown between tests, and the ability to execute dozens of tests in parallel.

When launched for a series of tests, the \texttt{Fuzzer} parses the Fuzz Specification, reads in the parameter vector, and passes it to the \texttt{Executor}, which is a multi-threaded system that coordinates fuzz test execution. The \texttt{Executor} serializes the specification into {mavros} \cite{mavros} messages mimicking real‐world remote-control (RC) stick inputs. By publishing RC‐style channel overrides and waypoint commands over MQTT into the autopilot container, the system exercises a similar command interface to that used in the field with physical flight controllers. The \texttt{Executor} dynamically injects specification-prescribed environmental conditions such as Wind or GPS perturbances, compass interference, and IMU (inertial measurement unit) noise into the simulation environment. It is also responsible for controlling other fuzzing variables, such as timing delays. 
The \texttt{Analyzer} collects and parses raw logs following each test execution to extract key diagnostic metrics such as attitude and path-tracking deviations, failsafe activations, mission completion status, exception flags, and other relevant flight data used for anomaly detection. Finally, upon completion of the mission or once mission failure is detected, the \texttt{Executor} tears down all containers and resets Gazebo to the base world file, restoring the environment to its default initial state.
After each Fuzz Specification is executed, results from all executed tests are transformed into truth tables, and the \texttt{Analyzer} identifies minimum-cut sets that induce failures.

\subsection{Applying the \fwnamebold Process}
To apply the \fwname process, we followed the previously outlined steps (cf.~\citefig{overview}). In Step 1 of Phase 1, one member of the research team with domain knowledge of \droneresponse and 8 years of experience working with sUAS, performed an initial hazard analysis producing the hazard tree depicted in~\citefig{hazards}. This was not intended to be exhaustive, and was guided by prior incidents revealing common types of errors reported in the literature~\cite{vierhauser2021hazard,di2023automated,wackwitz2015safety,plioutsias2018hazard}. 
In Step 2, three members of our research team (all co-authors of this paper) created the three  Fuzz Specifications depicted in Table~\ref{tab:fuzz-scenarios} as \fsc 1-3. The first Fuzz Specification was described earlier (cf. Listing~\ref{lst:fuzz-scenario-1}), and two additional specifications are described in the supplemental material.
The first specification (\fsc-1) tests simple mode transitions triggered by human actions during autonomous flight and is directly related to hazard \emph{H1}, and also to \emph{H3} for human triggered mode changes. The second (\fsc-2) tests failsafe transitions (related to \emph{H2}), while the third (\fsc-3) tests geofence interactions with human inputs, representing a feature interaction hazard (\emph{H4}).
In parallel to constructing Fuzz Specifications, two co-authors created three different mission specifications (Step 3), providing state/mode coverage for each of the Fuzz Specifications. The decision tree (Step 4), depicted in~\citefig{decision-tree}, was constructed based on a combination of our own domain knowledge and PX4 documentation \cite{px4modes}. It was initially constructed in the Summer of 2024, and has been iteratively evolved over the past year as part of early experimentation, using the Design Science approach. Constructing it took approximately 6 hours of initial effort, plus 1-2 hours of additional effort for revisions identified when \fwname produced false positives.

Next, in Phase 2, we executed the automated part of the pipeline on our SuT. Tests were generated for the three Fuzz Specifications using our \fwname  prototype. We generated 3,600 tests for \fsc-1, 6,480 for \fsc-2, and 1,080 for \fsc-3, identifying 10, 56, and 11 runs that ended in a FAILURE state, respectively. Running time on Ubuntu 22.04.3 LTS with i9-11900 processor, 4.5 TB SSD, 8 cores, 2.50GHz base, 64.0 GiB RAM took a total of 248 hours for all three Fuzz Specifications.  Steps 6 and 7 were then applied to identify failure cases, generate Fault Trees, and reduce each of them to a minimal cut-set representing the smallest conjunction of predicates guaranteeing failure. This produced 11 failures (cf.~Table \ref{tab:fault_analysis}), each generated as a visual Fault Tree. Three of these are shown in~\citefig{failure_cases},  with the complete set available in our supplemental material.

\vspace{-6pt}
\section{Results and Analysis}
\label{sec:application}

We now systematically report on the results of applying \fwname to \droneresponse and address each of the research questions in turn.

\subsection{RQ1 -- Effectiveness of \fwnamebold Process}
RQ1 investigates the extent to which \fwname identifies genuine behavioral faults. We measure this primarily by assessing the precision of the identified faults, and additionally by categorizing them by type. Results are reported in Table~\ref{tab:fault_analysis} and show that \fwname returned 11 cases. To analyze their correctness and to explore the underlying problems, the first author of this paper conducted two separate meetings with the \droneresponse lead Software Architect, a full-time professional Software Engineer with 8 years of experience working on various stages of \droneresponse. During these meetings, they inspected each fault tree and investigated both the \droneresponse generated logs and the PX4 flight control logs, to determine whether the candidate failure identified by \fwname was a {\it True Positive} or {\it False Positive}. 

To categorize each confirmed failure, three team members then applied a bottom-up approach whereby they discussed each failure case in depth, and assigned preliminary tags to characterize the nature of the issue. These tags served as a starting point for proposing categories, which were then refined collaboratively, merging overlapping terms, renaming for clarity, and converging on a small set of consistently applicable labels. Finally, for each category of failure, mitigations were identified.

{Results} are reported in \citetable{fault_analysis} and show that of the 11 faults, seven were categorized as correctly identified mode/state related failures (F1-F7), to be validated in the real-world tests, and one was classified as a valid fault associated with the PX4 autopilot code, and not directly impacted by mode and state transitions of the SuT (F8).
Additionally, we determined that faults (F9-11) were false positives. In the spirit of the iterative Design Science process, which we adopted throughout this process, these false positives were rectified by updating the node in the decision tree labeled ``Mode Change to LOITER'' to acknowledge that \mdstyle{AUTO.LOITER}, \mdstyle{POSCTL}, and throttle toggling all result in \mdstyle{POSCTL} in PX4. This class of false positives will therefore not be raised again in future fuzz tests. The identified mitigations are reported in the supplemental materials.%\vspace{-4pt}

\begin{reqbox}{\small Findings  RQ1: \fwname Automation Support}{}
\small
\fwname successfully identified seven failure cases relevant to state/mode transitions in the SuT. It also identified one failure related to the autopilot. Three false positive tests were caused by missing logic in the decision tree.
\end{reqbox}

\subsection{RQ2 -- Failure Identification}
We evaluate RQ2 by comparing the mode- and state-related transition errors identified as part of the normal testing process by the \droneresponse development team  against those detected by \mbox{\fwname}. The details of this comparison are as follows. In January 2024, we branched the then current {\it stable} codebase and created a new, frozen branch named {\it fuzz\_test}. Over the subsequent 18 months, development continued independently on a series of feature branches, with all changes ultimately merged back into {\it stable}, culminating in the July 2025 version. Therefore, in this experiment we compared the failures identified by \fwname in the frozen {\it fuzz\_test} baseline against those identified by the development team as the code evolved through to the July 2025 {\it stable} release. Notably, we ran \fwname tests against {\it fuzz\_test} in July 2025 and our findings had no impact upon the normal development cycle up until that time.
Further, since \fwname operated on the {\it fuzz\_test} branch, and the true set of failures was neither known nor knowable a priori, our comparison focused on (1) whether \fwname was able to detect all failures identified by the development team (i.e., recall), and (2) whether any additional failures were detected by \fwname that were not detected by the development team.

By July 16th, 2025 the {\it stable} branch was 889 commits ahead of {\it fuzz\_test}. Therefore, we first retrieved these commits from the \droneresponse repository, and then used a python parser to select the ones that referenced either an autopilot mode name or one of the 28 \droneresponse application-level state names. This query returned 147 commits. We then systematically inspected these commits and identified four that represented fixes for actual mode/state transition errors. In addition, we retrieved all issues that were currently open on January 24th, 2024, or were created between January 24th, 2024 and July 16th, 2025. From these we identified two relevant issues and four key commits related to state/mode transitions.

\begin{figure}[]
   
       \fbox{\includegraphics[width=.98\linewidth]{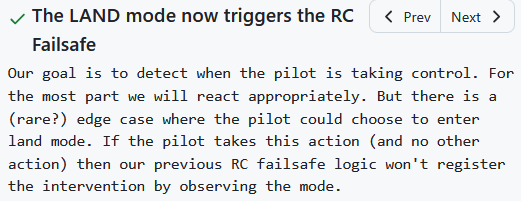}}
        
        \caption{Failure F1 identified by \fwname in the {\it safus} branch. It was found by developers in {\it stable}  and resolved in July 2024 via four commits that added the missing LAND failsafe.}
        \label{fig:fixed}
        \vspace{-16pt}
\end{figure}

We then provided the \droneresponse Software Architect with a list of the failures identified by \fwname, as well as the relevant commits and issues, and asked him to use this information plus his own knowledge of the project, to determine whether any failures found by \fwname had been independently found and addressed by the dev team. Of the two relevant issues, the Software Architect corroborated the first one as a relevant bug fix but explained that the second was related to a bug associated with a new state ({\it ReturnToCharge}), which had not been present in the original {\it fuzz\_test} branch. He did not identify any additional state/mode related failures that were fixed or observed during the 18 month period.

These results indicate that the development team only discovered one of the eight bugs (see Fig~\ref{fig:fixed}) discovered by \fwname. This led to three candidate interpretations: (1) the unfixed failure cases were non-critical and therefore it was inconsequential that the development team did not find them, (2) the testing process was inadequate and missed the failure cases, and/or (3) \fwname effectively revealed a unique class of failure not detected by the normal testing process. We reject (1), as the SuT is a life-critical search-and-rescue system, where failures occurring during deployment, could cause drones to land unexpectedly, fly away, or become stuck in the air—creating both safety and mission risks. With respect to (2), the testing process was robust by conventional standards, including automated unit tests, extensive simulation, and frequent field testing, yet it clearly underperformed in detecting the class of transition failures identified by \fwname. We therefore conclude that (3) is most plausible, and that integrating \fwname into the existing testing workflow identified additional, safety-relevant state–mode transition faults that were not exposed through standard validation.

%While this suggests that the failures represented potential edge cases, they introduced safety concerns where a failsafe mechanism may fail or work in an unexpected way, exactly when needed to address an emergent problem. Conversely, and perhaps more importantly, the single failure found by the development team was also successfully found by \fwname. \vspace{1pt}

\begin{table}[t]
    \centering
    \small
    \caption{The eight top ranked fault categories were analyzed across two interview sessions to determine root causes and to categorize as (i) A true positive mode/state related failure (\CIRCLE), (ii) a false positive failure (\Circle), or (iii) a valid failure but not directly associated to mode/state transitions (\RIGHTcircle).}
    \label{tab:fault_analysis}
\addtolength{\tabcolsep}{-2pt}
\renewcommand{\arraystretch}{1.0}
\begin{tabular}{|l|L{3cm}|L{4.0cm}|>{\centering\arraybackslash}p{0.3cm}|}
\hline
\bf{ID} & \bf{Category} & \bf{Description} & \textbf{\cmark} \\
\hline
\failid{F1} & \multirow{2}{*}{\parbox{3cm}{\vspace{0.6em}{\raggedright Mode change ignored from multiple states}}}
 & Land command ignored in \mdstyle{HOVER}. & \CIRCLE \\
\cline{1-1}\cline{3-4}
\failid{F2} &  & Human Control ignored during \mdstyle{TAKEOFF}. & \CIRCLE \\
\specialrule{.12em}{0pt}{0pt}
\failid{F3} & Mode Change Command causes thrashing & When \mdstyle{OFFBOARD} was activated during LAND, thrashing was observed between states. When  \mdstyle{OFFBOARD} reactivate it used an old setpoint, causing jerky flight. & \CIRCLE \\
\specialrule{.12em}{0pt}{0pt}
\failid{F4} & Delayed Mode Change & \mdstyle{POSCTL} not acknowledged during \mdstyle{RTL} triggered by geofence breach, until after LAND completed. & \CIRCLE \\
\specialrule{.12em}{0pt}{0pt}
\failid{F5} & Unclear requirements & \mdstyle{RTL} ignored during \mdstyle{TAKEOFF}. Treated as missing requirement as \mdstyle{RTL} should be handled as \mdstyle{LAND} during takeoff. & \CIRCLE \\
\specialrule{.12em}{0pt}{0pt}
\failid{F6} & Erratic mode changes caused by interference & GPS Noise created interference causing thrashing between \mdstyle{LAND} and \mdstyle{TAKEOFF} modes. & \CIRCLE \\
\specialrule{.12em}{0pt}{0pt}
\failid{F7} & Mode change ignored during failed state transition & Geofence breached with WARN action and \mdstyle{POSCTL} activated. \mdstyle{POSCTL} command ignored.& \CIRCLE\\
\specialrule{.12em}{0pt}{0pt}
\failid{F8} & PX4 issue within mode & PX4 Issue: Failure to disarm upon landing in \mdstyle{STABILIZED} mode. & \RIGHTcircle \\
\specialrule{.12em}{0pt}{0pt}
\failid{F9} & \multirow{3}{*}{\parbox{3cm}{\vspace{0.6em}{\raggedright Missing logic in Decision Tree. Updated to handle \mdstyle{AUTO.LOITER}~\& throttle toggle correctly in future tests}}}& Failure to recognize that Throttle toggling triggers \mdstyle{POSCTL}. & \Circle\\
\cline{1-1}\cline{3-4}
\failid{F10} &  & Failure to recognize that \mdstyle{AUTO.LOITER} is handled as \mdstyle{POSCTL} in RotorCraft during Flying state. & \Circle \\
\cline{1-1}\cline{3-4}
\failid{F11} &  & Failure to recognize that \mdstyle{AUTO.LOITER} is handled as \mdstyle{POSCTL} in RotorCraft during Landing state.& \Circle \\
\hline
\end{tabular}
\vspace{-12pt}
\end{table}

% \begin{figure}
%     \centering
%     % Row 1
%     \begin{subfigure}{0.47\textwidth}
%     \centering
%        \fbox{\includegraphics[width=.98\linewidth]{figures/issue-1.png}}
        
%         \caption{Failure F1 identified by \fwname, resolved in July 2024 via four commits to the {\it stable} branch addressing the missing LAND failsafe.}
%         \label{fig:fixed}
    
%     \end{subfigure}
%     \hfill
%     \begin{subfigure}{0.47\textwidth}
%         \centering
%     \fbox{\includegraphics[width=.98\linewidth]{figures/issue-2.png}}

%         \caption{New feature on {\it stable} branch, absent from the {\it safus} branch, and therefore not discoverable by \fwname.}

%     \end{subfigure}
%     \caption{Two mode-related issues from the stable \droneresponse branch: one correctly identified by \fwname as Failure F1, and one new feature not present in the {\it safus} branch, and thus not discoverable. However, the introduction of a new highly relevant feature, illustrates the need for \fwname' use as a regression testing instrument.}
    
% \end{figure}

%% we need to answer RQ1 here..
%\vspace{-4pt}
\begin{reqbox}{\small Findings RQ2: Alignment with Failures detected by the Dev Team}{}
\small
 \fwname identified seven types of failures related to state/mode transitions. During an 18-month time period, the development team, despite thousands of hours of simulation and hundreds of real-world flights, detected only one of these failures. The only additional failure related to state transitions that was detected by the team was due to a new feature, and therefore not present in the  {\it fuzz\_test} branch. 
\end{reqbox}

\subsection{RQ3 -- Field Test Validation}
To address our final research question, we validated our findings, where safe to do so, with physical sUAS using the {\it fuzz\_test} branch of our code and the same version of PX4 used in simulation. We created a representative field test for each identified failure, modifying only the flight coordinates to match the location of our outdoor site. We validated tests F1-F5 and F7. We excluded F6 because we could not easily control GPS factors, such as satellite geometry, in the real world, and excluded F8 because landing in stabilized mode is inadvisable as it could lead to a crash. Finally, we did not validate the three False Positives (F9-F11).

\newcolumntype{P}[1]{>{\raggedright\arraybackslash}p{#1}} % left-aligned p-column

\begin{table}[t]
\caption{Field test results executed on Physical sUAS running PX4. Each test shows the outcome and specifies whether it confirms the simulated results (\cmark) or not (\xmark).}
\vspace{-5pt}
\label{tab:field_results_px4}
\centering
\small
\setlength{\tabcolsep}{4pt}
\renewcommand{\arraystretch}{1.12}
\begin{tabular}{|P{.3cm}|L{7.15cm}|>{\centering\arraybackslash}p{0.3cm}|}
\hline
%& {\bf Test Outcome (PX4)}& \\ \hline
 \multicolumn{3}{|l|}{{\bf Test Outcome (PX4)}} \\ \hline

F1 & 
Land command ignored during hover. The drone initiated a landing but did not complete it, then proceeded to the next waypoint (offboard likely remained active). &\textbf{\cmark}\\ \hline

F2 & 
The system failed to acknowledge \mdstyle{POSCTL} and took off. Does not hand over to human for control. &\textbf{\cmark}\\ \hline

F3 & 
The drone nearly crashed after thrashing between offboard and land. Our testing pilot had to manually intervene to save the drone. &\textbf{\cmark}\\ \hline

F4 & 
Behavior not observed in the field, likely a simulation-to-physical flight delta. However, we noticed that geofence breaches behaved inconsistently in the field with PX4 and often breached earlier than expected. &\xmark\\ \hline

F5 & 
The system failed to acknowledge \mdstyle{RTL}. After shutting down the state machine, the vehicle could not be disarmed until we performed another manual takeoff. &\cmark\\ \hline

F7 & 
The sUAS immediately responded to \mdstyle{POSCTL}. &\textbf{\xmark}\\ \hline

\end{tabular}
\end{table}

The physical test setup included a hexacopter equipped with PX4 and a Jetson Xavier NX computation unit connected to a Ground Control Station via MeshRadio. Prior to each test, we configured  parameters, such as geofence actions, using QGroundControl~\cite{qground}, and sent the mission specification as a JSON file from \droneresponse's Ground Control Station to its onboard autonomous pilot. Each test involved two members of our team designated as a Computer Operator and a Remote Pilot in Command (RPIC). The Computer Operator was responsible for sending missions and monitoring current states and modes in a Graphical User Interface (GUI), while the RPIC was responsible for physically observing the sUAS.  Further, during test execution, the Computer Operator notified the RPIC when the targeted test state had been reached, and the RPIC then issued the designated MODE update using the RC transmitter. Both researchers visually observed the system’s behavior and recorded flight logs for post-test analysis. Table~\ref{tab:field_results_px4} summarizes the physical tests and their observed outcomes.

As reported in Table \ref{tab:field_results_px4}, results indicated strong, though not perfect, alignment between simulation and field outcomes. Four of the six failures reproduced the same behaviors observed in simulation. The \mdstyle{LAND} command was ignored during hover in F1; manual mode changes during takeoff were unacknowledged in F2; severe mode thrashing between \mdstyle{OFFBOARD} and \mdstyle{LAND} requiring manual recovery was observed in F3, and  the return-to-launch request and accompanying failsafe were both ignored in F5. However, tests F4 and F7 both exhibited simulation-to-field discrepancies related to geofence and failsafe handling. 

In summary, four out of six failure cases detected by \fwname were replicated in the physical world tests. Both of the tests that were not replicated included geofence mechanisms. In the case of F4, the simulation failed to acknowledge a POSCTL command issued immediately after a geofence triggered RTL until after the sUAS landed; while in the case of F7, a POSCTL command issued after a geofence WARN was not acknowledged. When tested in the field, both tests executed exactly as intended. These simulation failures suggest low fidelity of the simulated geofence functionality. %\vspace{-5pt}

\begin{reqbox}{\small Findings RQ3: \fwname Real-World Testing}{}
\small
Field validation showed that \fwname accurately reproduced 4 of 6 faults observed in simulation, with the remaining two representing simulation-to-reality deltas related to geofence and failsafe behavior. These results confirm that \fwname’s findings translate to real-world deployments while also exposing fidelity limits in current simulation environments.
\end{reqbox}

\subsection{Discussion}
These results demonstrate that \fwname is capable of uncovering meaningful and realistic faults in sUAS autonomy stacks, including issues that otherwise persisted for months despite ongoing development and testing. The automation of test oracle construction, grounded in state-machine and mode-transition reasoning, enabled the discovery of subtle failures such as inconsistent handling of mode change commands, unclear requirements, and simulation artifacts that obscured real-world behavior. Importantly, field validation confirmed that several of these faults manifested in physical deployments and were not previously identified through standard testing pipelines. At the same time, discrepancies observed in F4 and F7 highlight limitations in simulator fidelity especially with respect to geofence handling and failsafe actions.  As geofences provide essential safety constraints, this clearly indicates the need for higher-fidelity geofence models that properly reflect the interactions between the autopilot and its environment~\cite{simgap2}.~Nevertheless, both test results are valuable. Tests that confirm simulated behavior in the real-world provide confidence that fixes tested in simulation will also hold in the physical world; while inconsistencies in the sim-to-real progression bring awareness for parts of the system where simulation results cannot be trusted and improvements in the underlying simulation platform are needed.  While our experiments focused on a small number of targeted Fuzz Specifications, they demonstrated \fwname's ability to reveal failure cases missed by conventional simulation and field testing.

% \begin{figure}
%     \centering
%     \includegraphics[width=\linewidth]{figures/placeholder.jpg}
%     \caption{Timeline of F7 field test. Even though we are short of space, I think we need this.}
%     \label{fig:f7}
% \end{figure}  
\section{Threats to Validity}
\label{sec:threats}
The work presented in this paper is subject to several limitations that may affect generalizability and interpretability of the results.
First, the evaluation was conducted using a single SuT, specifically the \droneresponse multi-sUAS system. While this allowed for an in-depth analysis of the functionality, code, and failures, the results may not fully translate to other sUAS systems with different architectures, state machines, or flight controllers used. Second, while simulation-to-reality deltas were observed in tests associated with the geofence, these highlight potential fidelity limits in high-end simulators such as Gazebo. Reproducing four out of the six failures in physical flights validates the utility of our approach. At the same time, sim-to-real discrepancies point to open challenges in accurately modeling environmental and controller dynamics. In future work, we plan to extend our framework to better characterize and reduce these gaps.

Third, the study did not include a structured user evaluation. Although the visualization and fault categorization outputs were shared with \droneresponse developers, no formal user study was conducted to assess usability, interpretability, or decision support effectiveness. However, anecdotal evidence from the in-depth interviews with the \droneresponse lead architect strongly suggests that the visualized Fault Trees provide value for analyzing failures.

Fourth, we have not compared \fwname to a baseline approach beyond RQ2, which compared its outcomes against failures detected by the development team and field-testers. The \droneresponse team follows a robust devops approach; however, \fwname clearly detected additional failures. We also did not compare against a more formal approach, primarily because of the complexity of the application-level and lower-level state machines, and its continual evolution. In contrast, the fuzz-testing approach we have presented can easily be extended to cover new functionality, simply by defining new Fuzz Specifications. 
Our approach was designed to accommodate real-world development constraints, but this could limit the applicability to other domains/types of systems where specification-based and/or formal verification is mandated. 
Despite these limitations, the findings provide practical and valuable insights into the challenges of autonomous multi-level mode transitions and the utility of lightweight analysis methods in a multi-sUAS system.

\section{Related Work}
\label{sec:relwork}
% (Link: https://www.dropbox.com/scl/fo/cr17qklv559g5uihgqxfz/ANRS8vOABmhXmCqUZvpHPQ4?rlkey=4mspwaj5n8a2tfrmwlsufovu0&st=5pt7ghbk&dl=0)

We focus related work on three relevant areas of general \emph{CPS and sUAS testing}~\cite{abbaspour2015survey}, \emph{fuzz testing}~\cite{fuzz-survey:2022}, and \emph{safety}. \vspace{2pt}

\textbf{Testing of CPS:}
CPS testing incorporates diverse facets including hardware testing, testing of extra-functional properties, as well as integration and system testing~\cite{abbaspour2015survey,zhou2018review}. De Liso and Wen~\cite{de2024camba} presented CAMBA, a cost-aware, mutation-based test case generation algorithm for UAVs. Their work focused on a smart obstacle-placement system to test safe flight behavior. \cite{mandrioli2025testing} combined control-theoretical design assumptions with metamorphic testing and genetic programming.  Instead of relying on requirements and input traces, they defined metamorphic relations across inputs and outputs of multiple test cases. 
Liang~\etal[garl2025] presented the GARL framework, combining a genetic algorithm and reinforcement learning to generate landing violation cases for sUAS landing systems. Like us, they combined simulation-based and real-world tests for diverse landing scenarios. However, \fwname addresses a far broader scope of mission types and includes human interactions.
Duvvuru~\etal[ankit2025ICSE] introduced AutoSimTest, a framework using LLM agents to automate simulation-testing of sUAS. Similar to our work, they generate test scenarios and simulation configurations, but lack structured analysis support as we do with our Fault Trees.
Many other testing techniques have been applied to sUAS, including vision-based testing~\cite{bu2015general}, and data-driven approaches~\cite {sarkar2020pie}. However, they are typically limited to narrow aspects of a CPS, neglecting human-CPS-interaction, covering only a limited space (e.g., security~\cite{hagerman2016security}), with tests often limited to simulations. \vspace{2pt}

\textbf{Fuzzing:} Other researchers have proposed fuzz testing for robotic applications.
Delgado~\etal[delgado2021fuzz] presented a fuzzer for ROS-based systems using SMACH (a library for plan execution), where fuzzing is performed on the SMACH states. \droneresponse also uses SMACH to support its application-level state machine. 
Woodlief~\etal~\cite{woodlief2021fuzzing} developed PHYS-FUZZ for fuzzing physical attributes, such as trajectories. RoboFuzz~\cite{robofuzz:2022}, designed for integration with ROS as a feedback-driven fuzzing framework, has also been applied to PX4 Quadcopter drones. Wang~\etal[wang2025dpfuzzer] proposed DPFuzzer, an automated framework for detecting vulnerabilities in drone path planners. Like us, they generated diverse
 scenarios using fuzzing techniques. However, while all these approaches use fuzzing, they primarily focus on flight controller properties, lack support for human-interaction-based fuzzing, and do not incorporate subsequent safety analysis.
In contrast, our own prior work applied fuzzing within the sUAS domain~\cite{chambers2024hifuzz}, incorporating human–interaction failures with a staged progression from proxy-human simulation to human-in-the-loop and safety-aware field tests. However, it provided limited support for diagnosing the root cause of observed failures. This limitation motivated \fwname, which focuses on test automation using a decision-tree based failure oracle and automated diagnostic analyses that include Fault Tree generation and visualization. \fwname further incorporates substantial timing mutations, which can trigger race conditions during state transitions, as well as realistic environmental perturbations such as compass interference, thereby enabling high-throughput testing with explainable failures.\vspace{2pt}

\textbf{CPS Safety Analysis \& Assurance:} FT-MOEA, by  Jimenez-Roa~\emph{et al.}~\cite{jimenez2022automatic}  leverages multi-objective evolutionary algorithms to automatically recover Fault Trees from system data, easing manual and time-consuming Fault Tree Analysis. Like them we use the generated Fault Trees to aid project stakeholders in investigating errors and identifying root causes. 
Focusing on formal verification for CPS, Heitmeyer and Leonard~\cite{heitmeyer2015obtaining} present FORMAL, supporting formal modeling and symbolic execution of CPS.
Safety assurance cases are widely used in safety-critical domains, and requiring their use for sUAS systems is an active research area with an active research community. Most notably, Denney and Pai~\cite{denney2012lightweight}  studied modular safety cases, facilitating the capture and maintenance of safety-related sUAS behavior. Similarly, as part of our our previous work, we focused on ``interlocking'' Safety Assurance Cases (SACs), combining infrastructure-specific and sUAS-specific aspects into a safety argument~\cite{vierhauser2019interlocking}.
Kreutz~\etal[kreutz2025modeling] presented a method for modeling adaptation spaces using Contextual Safety Concept Trees for robotic systems. They formalized dependencies as fuzzy inference systems~\cite{cherkassky1998fuzzy}, and used them to evaluate safety requirements at runtime. We also use minimal cut sets of Fault Trees, but focus upon providing humans with support for root cause analysis.
While these approaches contribute to sUAS safety, their focus is primarily on manually created SACs, and does not include their automated creation or use in the testing process.

\section{Conclusions}
\label{sec:conclusions}

In this paper, we have introduced \fwname,  a novel fuzzing pipeline to validate the behavior of sUAS across multiple layers of control logic, including application-level state machines, flight controller modes, failsafes, and human interactions. By generating realistic and semantically meaningful test scenarios incorporating varying timing conditions and environmental factors,  \fwname detects transition failures and hazardous interactions that originate from both simple and complex system interactions. Based on this, dynamically generated Fault Trees support stakeholders in diagnosing root causes and improving system resilience. We have validated our approach through a series of high-fidelity simulations and real-world field tests.
The findings are of potential value to both practitioners and researchers. 
From a practitioner perspective, \fwname enhances existing development and testing processes by identifying transition-related faults, timing hazards, and unexpected behavior sequences that are often inaccessible through manual testing or ad hoc flight evaluations. 
From a research perspective, \fwname provides a structured way to study transition-centric failure modes, cross-layer hazards that have historically contributed to many accidents but remain under-examined in existing verification research.

Future work will investigate the applicability of \fwname to a broader range of tests generated in simulation and corroborated through physical testing, with particular emphasis on complex mode transitions, control-handoff behaviors, and interactions among application-level and flight-controller state machines.  
In addition, we plan to conduct targeted user studies with developers and testers to assess how effectively \fwname’s diagnostic outputs support fault understanding, debugging efficiency, and confidence in testing outcomes.

\section{Acknowledgments}
Work in this paper was primarily funded by the USA National Science Foundation under Grant \# 1931962.

\bibliographystyle{ACM-Reference-Format}
\balance
\bibliography{ICSE2026}

@inproceedings{chambers2024hifuzz,
  title = {{HIFuzz: Human Interaction Fuzzing for Small Unmanned Aerial Vehicles}},
  author = {Chambers, Theodore and Vierhauser, Michael and Agrawal, Ankit and Murphy, Michael and Brauer, Jason Matthew and Purandare, Salil and Cohen, Myra B. and Cleland-Huang, Jane},
  booktitle = {Proc. of the CHI Conference on Human Factors in Computing Systems},
  pages = {1--14},
  year = {2024},
  address = {Honolulu, HI, USA},
  month = may,
  organization = {ACM},
  isbn = {979-8-4007-0330-0}
}

@online{Hambling2020GPSInterferenceDroneCrash,
  author       = {David Hambling},
  title        = {Drone Crash Due To {GPS} Interference in U.K. Raises Safety Questions},
  year         = {2020},
  month        = aug,
  date         = {2020-08-10},
  organization = {Forbes},
  url          = {https://www.forbes.com/sites/davidhambling/2020/08/10/investigation-finds-gps-interference-caused-uk-survey-drone-crash/},
  note         = {Editors' Pick; Business – Aerospace \& Defense},
}

@article{DBLP:journals/corr/abs-2505-08060,
  author       = {Pedro Alarcon Granadeno and
                  Jane Cleland{-}Huang},
  title        = {Land-Coverage Aware Path-Planning for Multi-UAV Swarms in Search and
                  Rescue Scenarios},
  journal      = {CoRR},
  volume       = {abs/2505.08060},
  year         = {2025},
  url          = {https://doi.org/10.48550/arXiv.2505.08060},
  doi          = {10.48550/ARXIV.2505.08060},
  eprinttype    = {arXiv},
  eprint       = {2505.08060},
  bibsource    = {dblp computer science bibliography, https://dblp.org}
}

@inproceedings{Chambers2025UTMOnEntry,
  title        = {Automated On-Entry Decision-Making for {UTM} Zones Based on Reputations and Certifications},
  author       = {Chambers, Theodore P. and Granadeno, Pedro and Gohar, Usman and Hunter, Michael C. and Russell Bernal, Arturo Miguel and Tang, Wenyi and Al Islam, Md Nafee and Cohen, Myra and Jung, Taeho and Lutz, Robyn and Cleland-Huang, Jane},
  booktitle    = {AIAA Aviation Forum and ASCEND 2025},
  year         = {2025},
  pages        = {3567}
}

@article{DBLP:journals/taas/ClelandHuangCZCAV24,
  author       = {Jane Cleland{-}Huang and
                  Theodore Chambers and
                  Sebasti{\'{a}}n Zudaire and
                  Muhammed Tawfiq Chowdhury and
                  Ankit Agrawal and
                  Michael Vierhauser},
  title        = {Human-machine Teaming with Small Unmanned Aerial Systems in a {MAPE-K}
                  Environment},
  journal      = {{ACM} Trans. Auton. Adapt. Syst.},
  volume       = {19},
  number       = {1},
  pages        = {3:1--3:35},
  year         = {2024},
  url          = {https://doi.org/10.1145/3618001},
  doi          = {10.1145/3618001},
  bibsource    = {dblp computer science bibliography, https://dblp.org}
}

@book{wieringa2014design,
  author    = {Wieringa, Roel J.},
  title     = {Design Science Methodology for Information Systems and Software Engineering},
  publisher = {Springer},
  address   = {London, Germany},
  year      = {2014},
  isbn      = {978-3-662-43838-1},
  doi       = {10.1007/978-3-662-43839-8},
}

@article{anto2023,
author = {Anto, Kelvin and Swain, AK and Roop, Partha},
year = {2023},
month = {01},
pages = {1-1},
title = {A Novel Framework for the Design of Resilient Cyber-Physical Systems Using Control Theory and Formal Methods},
volume = {PP},
journal = {IEEE Access},
doi = {10.1109/ACCESS.2023.3295421}
}

@inproceedings{DBLP:conf/seams/PurandareSICC23,
  author       = {Salil Purandare and
                  Urjoshi Sinha and
                  Md Nafee Al Islam and
                  Jane Cleland{-}Huang and
                  Myra B. Cohen},
  title        = {Self-Adaptive Mechanisms for Misconfigurations in Small Uncrewed Aerial
                  Systems},
  booktitle    = {Proc. of the 18th {IEEE/ACM} Symposium on Software Engineering for Adaptive and Self-Managing Systems},
  pages        = {169--180},
publisher = {IEEE Computer Society},
address = {Los Alamitos, CA, USA},
  year         = {2023}
}

@misc{skydance_incident,
  title = {Ex-Skydance Exec Piloted Drone That Crashed Into Firefighting Helicopter},
  author = {Gardner, Chris},
  year = {2022},
  howpublished = {https://www.hollywoodreporter.com/news/local-news/ex-skydance-exec-piloted-drone-crashed-plane-palisades-fire-1236123911},
  note = {[Last accessed 01-12-2025]}
}

@article{cui2020elbow,
  title={Introduction to the k-means clustering algorithm based on the elbow method},
  author={Cui, Mengyao and others},
  journal={Accounting, Auditing and Finance},
  volume={1},
  number={1},
  pages={5--8},
  year={2020},
  publisher={Clausius Scientific Press}
}

@article{quine1952problem,
 ISSN = {00029890, 19300972},
 URL = {http://www.jstor.org/stable/2308219},
 author = {W. V. Quine},
 journal = {The American Mathematical Monthly},
 number = {8},
 pages = {521--531},
 publisher = {[Taylor & Francis, Ltd., Mathematical Association of America]},
 title = {The Problem of Simplifying Truth Functions},
 urldate = {2025-07-15},
 volume = {59},
 year = {1952}
}

@inproceedings{abbaspour2015survey,
 author = {Abbaspour Asadollah, Sara and Inam, Rafia and Hansson, Hans},
 booktitle = {Proc. of the  27th IFIP WG 6.1 International Conference on Testing Software and Systems},
 publisher = {Springer International Publishing},
 pages = {194--207},
address={Cham},
 title = {A Survey on Testing for Cyber Physical System},
 year = {2015}
}

@article{Adams2009,
 author = {Julie A. Adams and Curtis M. Humphrey and Michael A. Goodrich and Joseph L. Cooper and Bryan S. Morse and Cameron Engh and Nathan Rasmussen},
 journal = {Journal of Cognitive Engineering and Decision Making},
 number = {1},
 pages = {1-26},
 title = {Cognitive Task Analysis for Developing Unmanned Aerial Vehicle Wilderness Search Support},
 volume = {3},
 year = {2009}
}

@misc{ardupilot,
 author = {ArduPilot},
 howpublished = {\url{http://ardupilot.org}},
 note = {[Last accessed 01-12-2025]},
 year = {2025}
}

@inproceedings{campusano2021towards,
 author = {Campusano, Miguel and Jensen, Kjeld and Schultz, Ulrik Pagh},
 booktitle = {Proc. of the 2021 IEEE/ACM 3rd Int'l Workshop on Robotics Software Engineering (RoSE)},
publisher = {IEEE Computer Society},
address = {Los Alamitos, CA, USA},
 pages = {63--66},
 title = {Towards a Service-Oriented U-Space Architecture for Autonomous Drone Operations},
 year = {2021}
}

@article{cleland2018dronology,
 author = {Cleland-Huang, Jane and Vierhauser, Michael and Bayley, Sean},
 journal = {arXiv preprint arXiv:1804.02423},
 title = {Dronology: An incubator for cyber-physical system research},
 year = {2018}
}

@inproceedings{DBLP:conf/chi/AgrawalABCFHHTK20,
 author = {Ankit Agrawal and
Sophia J. Abraham and
Benjamin Burger and
Chichi Christine and
Luke Fraser and
John M. Hoeksema and
Sarah Hwang and
Elizabeth Travnik and
Shreya Kumar and
Walter J. Scheirer and
Jane Cleland{-}Huang and
Michael Vierhauser and
Ryan Bauer and
Steve Cox},
 bibsource = {dblp computer science bibliography, https://dblp.org},
 booktitle = {{CHI} '20: {CHI} Conference on Human Factors in Computing Systems,
Honolulu, HI, USA, April 25-30, 2020},
 doi = {10.1145/3313831.3376825},
 editor = {Regina Bernhaupt and
Florian 'Floyd' Mueller and
David Verweij and
Josh Andres and
Joanna McGrenere and
Andy Cockburn and
Ignacio Avellino and
Alix Goguey and
Pernille Bj{\o}n and
Shengdong Zhao and
Briane Paul Samson and
Rafal Kocielnik},
 pages = {1--13},
 publisher = {{ACM}},
 title = {The Next Generation of Human-Drone Partnerships: Co-Designing an Emergency
Response System},
 url = {https://doi.org/10.1145/3313831.3376825},
 year = {2020}
}

@inproceedings{denney2012lightweight,
 author = {Denney, Ewen and Pai, Ganesh},
 booktitle = {Proc. of the International Conference on Computer Safety, Reliability, and Security},
 publisher = {Springer},
 pages = {1--12},
 title = {A lightweight methodology for safety case assembly},
 year = {2012}
}

@inproceedings{frasheri2018adaptive,
 author = {Frasheri, Mirgita and C{\"u}r{\"u}kl{\"u}, Baran and Esktr{\"o}m, Mikael and Papadopoulos, Alessandro Vittorio},
 booktitle = {Proc. of the 12th International Conference on Self-Adaptive and Self-Organizing Systems},
 publisher = {IEEE Computer Society},
address = {Los Alamitos, CA, USA},
 pages = {150--155},
 title = {Adaptive autonomy in a search and rescue scenario},
 year = {2018}
}

@article{fuzz-survey:2022,
 address = {New York, NY, USA},
 articleno = {230},
 author = {Zhu, Xiaogang and Wen, Sheng and Camtepe, Seyit and Xiang, Yang},
 doi = {10.1145/3512345},
 issn = {0360-0300},
 issue_date = {January 2022},
 journal = {ACM Comput. Surv.},
 month = {sep},
 number = {11s},
 numpages = {36},
 publisher = {Association for Computing Machinery},
 title = {Fuzzing: A Survey for Roadmap},
 url = {https://doi.org/10.1145/3512345},
 volume = {54},
 year = {2022}
}

@misc{gazebo,
 author = {{Open Robotics}},
 howpublished = {\url{https://gazebosim.org}},
 note = {[Last accessed 01-07-2025]},
 title = {{Gazebo}},
 year = {2025}
}

@misc{PX4,
 author = {{PX4 - Open Source Autopilot}},
 howpublished = {\url{https://px4.io}},
 note = {[Last accessed 01-12-2025]},
 title = {{PX4}},
 year = {2025}
}

@misc{mavlink,
 author = {{DroneCode}},
 howpublished = {\url{https://mavlink.io/en}},
 note = {[Last accessed 01-12-2025]},
 title = {{MAVLink - Developer Guide}},
year = {2025}
}

@misc{mavros,
 author = {Vladimir Ermakov},
 publisher = {GitHub},
 title = {{MAVROS}},
 howpublished = {https://github.com/mavlink/mavros},
 note = {[Last accessed 01-12-2025]},
year = {2025}
}

@inproceedings{mcaree2016model,
 author = {McAree, Owen and Aitken, Jonathan M and Veres, Sandor M},
 booktitle = {Proc. of the 11th International Conference on Control},
publisher = {IEEE Computer Society},
address = {Los Alamitos, CA, USA},
 pages = {1--6},
 title = {A model based design framework for safety verification of a semi-autonomous inspection drone},
 year = {2016}
}

@misc{px4modes,
 author = {{PX4}},
 howpublished = {\url{https://docs.px4.io/main/en/flight_modes_mc/}},
 note = {[Last accessed 01-12-2025]},
 title = {{Flight Controller Modes}},
 year = {2023}
}

@misc{qground,
 author = {{QGroundControl}},
 howpublished = {\url{http://qgroundcontrol.com}},
 note = {[Last accessed 01-07-2025]},
 title = {{Ground Control Station}},
 year = {2025}
}

@inproceedings{robofuzz:2022,
 address = {New York, NY, USA},
 author = {Kim, Seulbae and Kim, Taesoo},
 booktitle = {Proc. of the 30th ACM Joint European Software Engineering Conference and Symposium on the Foundations of Software Engineering},
 location = {Singapore, Singapore},
 numpages = {12},
 pages = {447–458},
 publisher = {Association for Computing Machinery},
 series = {ESEC/FSE 2022},
 title = {{RoboFuzz: Fuzzing Robotic Systems over Robot Operating System (ROS) for Finding Correctness Bugs}},
 year = {2022}
}

@inproceedings{vierhauser2021hazard,
 author = {Vierhauser, Michael and Islam, Md Nafee Al and Agrawal, Ankit and Cleland-Huang, Jane and Mason, James},
 booktitle = {Proc. of the 29th ACM Joint Meeting on European Software Engineering Conference and Symposium on the Foundations of Software Engineering},
 pages = {8--19},
 title = {Hazard analysis for human-on-the-loop interactions in sUAS systems},
publisher = {Association for Computing Machinery},
address = {New York, NY, USA},
 year = {2021}
}

@article{zhou2018review,
  title={Review on testing of cyber physical systems: Methods and testbeds},
  author={Zhou, Xin and Gou, Xiaodong and Huang, Tingting and Yang, Shunkun},
  journal={IEEE Access},
  volume={6},
  pages={52179--52194},
  year={2018},
  publisher={IEEE}
}

@inproceedings{simgap2,
  title={Leveraging multiple simulators for crossing the reality gap},
  author={Boeing, Adrian and Br{\"a}unl, Thomas},
  booktitle={Proc. of the 12th International Conference on Control Automation Robotics \& Vision},
  pages={1113--1119},
  year={2012},
  organization={IEEE}
}

@inproceedings{woodlief2021fuzzing,
  title={Fuzzing mobile robot environments for fast automated crash detection},
  author={Woodlief, Trey and Elbaum, Sebastian and Sullivan, Kevin},
  booktitle={2021 IEEE International Conference on Robotics and Automation (ICRA)},
  pages={5417--5423},
  year={2021},
  publisher = {IEEE Computer Society},
address = {Los Alamitos, CA, USA},
}

@article{vierhauser2019interlocking,
  title={Interlocking safety cases for unmanned autonomous systems in shared airspaces},
  author={Vierhauser, Michael and Bayley, Sean and Wyngaard, Jane and Xiong, Wandi and Cheng, Jinghui and Huseman, Joshua and Lutz, Robyn and Cleland-Huang, Jane},
  journal={IEEE transactions on software engineering},
  volume={47},
  number={5},
  pages={899--918},
  year={2019},
  publisher={IEEE}
}

@inproceedings{de2024camba,
  title={CAMBA CPS-UAV at the SBFT Tool Competition 2024: CAMBA: Cost-Aware Mutation-Based Test Case Generation for Unmanned Aerial Vehicles},
  author={De Liso, Marco and Soi, Zhi Wen},
  booktitle={Proc. of the 17th ACM/IEEE International Workshop on Search-Based and Fuzz Testing},
  pages={47--48},
publisher = {Association for Computing Machinery},
address = {New York, NY, USA},
  year={2024}
}

@article{mandrioli2025testing,
  title={Testing {CPS} with design assumptions-based metamorphic relations and genetic programming},
  author={Mandrioli, Claudio and Shin, Seung Yeob and Bianculli, Domenico and Briand, Lionel},
  journal={IEEE Transactions on Software Engineering},
  year={2025},
volume={51},
  number={6},
  publisher={IEEE}
}

@article{jimenez2022automatic,
  title={Automatic inference of fault tree models via multi-objective evolutionary algorithms},
  author={Jimenez-Roa, Lisandro Arturo and Heskes, Tom and Tinga, Tiedo and Stoelinga, Mari{\"e}lle},
  journal={IEEE transactions on dependable and secure computing},
  volume={20},
  number={4},
  pages={3317--3327},
  year={2022},
  publisher={IEEE}
}

@inproceedings{heitmeyer2015obtaining,
  title={Obtaining trust in autonomous systems: Tools for formal model synthesis and validation},
  author={Heitmeyer, Constance L and Leonard, Elizabeth I},
  booktitle={2015 IEEE/ACM 3rd FME Workshop on Formal Methods in Software Engineering},
  pages={54--60},
  year={2015},
  publisher = {IEEE Computer Society},
address = {Los Alamitos, CA, USA},
}

@inproceedings{bu2015general,
  title={General simulation platform for vision based UAV testing},
  author={Bu, Qing and Wan, Fuhua and Xie, Zhen and Ren, Qinhu and Zhang, Jianhua and Liu, Sheng},
  booktitle={2015 IEEE International Conference on Information and Automation},
  pages={2512--2516},
  year={2015},
  publisher = {IEEE Computer Society},
address = {Los Alamitos, CA, USA},
}

@article{sarkar2020pie,
  title={Pie: a tool for data-driven autonomous uav flight testing},
  author={Sarkar, Mrinmoy and Homaifar, Abdollah and Erol, Berat A and Behniapoor, Mohammadreza and Tunstel, Edward},
  journal={Journal of Intelligent \& Robotic Systems},
  volume={98},
  pages={421--438},
  year={2020},
  publisher={Springer}
}

@inproceedings{hagerman2016security,
  title={Security testing of an unmanned aerial vehicle {(UAV)}},
  author={Hagerman, Seana and Andrews, Anneliese and Oakes, Stephen},
  booktitle={Proc. of the 2016 Cybersecurity Symposium},
  pages={26--31},
  year={2016},
 publisher = {IEEE Computer Society},
address = {Los Alamitos, CA, USA},
}

@article{cherkassky1998fuzzy,
  title={Fuzzy inference systems: a critical review},
  author={Cherkassky, Vladimir},
  journal={Computational intelligence: soft computing and fuzzy-neuro integration with applications},
  pages={177--197},
  year={1998},
  publisher={Springer}
}

@inproceedings{rocamora2024behavior,
  title={A behavior tree approach for battery-aware inspection of large structures using drones},
  author={Rocamora, Bernardo Martinez and Simplicio, Paulo VG and Pereira, Guilherme AS},
  booktitle={2024 International Conference on Unmanned Aircraft Systems (ICUAS)},
  pages={234--240},
  year={2024},
publisher = {IEEE Computer Society},
address = {Los Alamitos, CA, USA},
}

@inproceedings{siewert2019fail,
  title={Fail-safe, fail-secure experiments for small UAS and UAM traffic in urban airspace},
  author={Siewert, Sam and Sampigethaya, Krishna and Buchholz, Jonathan and Rizor, Steve},
  booktitle={Proc. of the 2019 IEEE/AIAA 38th Digital Avionics Systems Conference},
  pages={1--7},
  year={2019},
  publisher = {IEEE Computer Society},
address = {Los Alamitos, CA, USA},
}

@incollection{falayi2025edge,
  title={Edge intelligence in smart transportation CPS},
  author={Falayi, Ayodeji and Wang, Qianlong and Yu, Wei},
  booktitle={Edge Intelligence in Cyber-Physical Systems},
  pages={193--219},
  year={2025},
  publisher={Elsevier}
}

@incollection{kumar2024implementation,
  title={Implementation of Intelligent CPS for Integrating the Industry and Manufacturing Process},
  author={Kumar, T Rajasanthosh and Kawade, Mahesh M and Bharti, Gaurav Kumar and Laxmaiah, G},
  booktitle={AI-Driven IoT Systems for Industry 4.0},
  pages={273--288},
  year={2024},
  publisher={CRC Press}
}

@article{haque2014review,
  title={Review of cyber-physical system in healthcare},
  author={Haque, Shah Ahsanul and Aziz, Syed Mahfuzul and Rahman, Mustafizur},
  journal={International Journal of Distributed Sensor Networks},
  volume={10},
  number={4},
  pages={217415},
  year={2014},
  publisher={SAGE Publications Sage UK: London, England}
}

@incollection{anand2023drones,
  title={Drones for disaster response and management},
  author={Anand, Jose and Aasish, C and Narayanan, S Syam and Ahmed, R Asad},
  booktitle={Internet of Drones},
  pages={177--200},
  year={2023},
  publisher={CRC Press}
}

@article{golabi2022towards,
  title={Towards automated hazard analysis for CPS security with application to CSTR system},
  author={Golabi, Arash and Erradi, Abdelkarim and Tantawy, Ashraf},
  journal={Journal of Process Control},
  volume={115},
  pages={100--111},
  year={2022},
  publisher={Elsevier}
}

@article{plioutsias2018hazard,
  title={Hazard analysis and safety requirements for small drone operations: to what extent do popular drones embed safety?},
  author={Plioutsias, Anastasios and Karanikas, Nektarios and Chatzimihailidou, Maria Mikela},
  journal={Risk Analysis},
  volume={38},
  number={3},
  pages={562--584},
  year={2018},
  publisher={Wiley Online Library}
}

@article{di2023automated,
  title={Automated identification and qualitative characterization of safety concerns reported in uav software platforms},
  author={Di Sorbo, Andrea and Zampetti, Fiorella and Visaggio, Aaron and Di Penta, Massimiliano and Panichella, Sebastiano},
  journal={ACM Transactions on Software Engineering and Methodology},
  volume={32},
  number={3},
  pages={1--37},
  year={2023},
  publisher={ACM New York, NY}
}

@article{wackwitz2015safety,
  title={Safety risk assessment for uav operation},
  author={Wackwitz, Kay and Boedecker, Hendrick},
  journal={Drone Industry Insights, Safe Airspace Integration Project, Part One, Hamburg, Germany},
  pages={31--53},
  number={},
volume={},
  year={2015}
}

@article{bartocci2021cpsdebug,
  title={CPSDebug: Automatic failure explanation in CPS models},
  author={Bartocci, Ezio and Manjunath, Niveditha and Mariani, Leonardo and Mateis, Cristinel and Ni{\v{c}}kovi{\'c}, Dejan},
  journal={International Journal on Software Tools for Technology Transfer},
  volume={23},
  number={5},
  pages={783--796},
  year={2021},
  publisher={Springer}
}

@inproceedings{amir2017hybrid,
  title={Hybrid state machine model for fast model predictive control: Application to path tracking},
  author={Amir, Maral and Givargis, Tony},
  booktitle={Proc. of the IEEE/ACM International Conference on Computer-Aided Design (Iccad)},
  pages={185--192},
  year={2017},
  publisher = {IEEE Computer Society},
address = {Los Alamitos, CA, USA},
}

@inproceedings{smyczynski2017autonomous,
  title={Autonomous drone control system for object tracking: Flexible system design with implementation example},
  author={Smyczy{\'n}ski, Pawe{\l} and Starzec, {\L}ukasz and Granosik, Grzegorz},
  booktitle={Proc. of the 22nd International Conference on Methods and Models in Automation and Robotics},
  pages={734--738},
  year={2017},
  publisher = {IEEE Computer Society},
address = {Los Alamitos, CA, USA},
}

@article{bolton2013using,
  title={Using formal verification to evaluate human-automation interaction: A review},
  author={Bolton, Matthew L and Bass, Ellen J and Siminiceanu, Radu I},
  journal={IEEE Transactions on Systems, Man, and Cybernetics: Systems},
  volume={43},
  number={3},
  pages={488--503},
  year={2013},
publisher = {IEEE Computer Society},
address = {Los Alamitos, CA, USA},
}

@inproceedings{lemieux2018fairfuzz,
title = {{FairFuzz}: a targeted mutation strategy for increasing greybox fuzz testing coverage},

  author={Lemieux, Caroline and Sen, Koushik},
  booktitle = {Proc. of the 33rd ACM/IEEE International Conference on Automated Software Engineering},
publisher = {Association for Computing Machinery},
address = {New York, NY, USA},
  pages={475--485},
  year={2018}
}

@inproceedings{visser2020coastal,
  title={Coastal: Combining concolic and fuzzing for Java (competition contribution)},
  author={Visser, Willem and Geldenhuys, Jaco},
  booktitle={Proc. of the International Conference on Tools and Algorithms for the Construction and Analysis of Systems},
  pages={373--377},
  year={2020},
  publisher={Springer},
address = {Cham},

}

@inproceedings{bekrar2011finding,
  title={Finding software vulnerabilities by smart fuzzing},
  author={Bekrar, Sofia and Bekrar, Chaouki and Groz, Roland and Mounier, Laurent},
  booktitle={Proc. of the 2011 Fourth IEEE International Conference on Software Testing, Verification and Validation},
  pages={427--430},
  year={2011},
 publisher = {IEEE Computer Society},
address = {Los Alamitos, CA, USA},
}

@inproceedings{holler2012fuzzing,
  title={Fuzzing with code fragments},
  author={Holler, Christian and Herzig, Kim and Zeller, Andreas},
  booktitle={Proc. of the 21st USENIX Security Symposium},
  pages={445--458},
  year={2012}
}

@misc{droneresponse,
  title = {Drone Response sUAS Platform},
  author = {Drone Response},
  year = {2025},
  howpublished = {https://droneresponse.ai},
  note = {[Last accessed 31-12-2025]}
}

\clearpage

\end{document}